\documentclass[aps,prb,onecolumn,preprint,superscriptaddress,showkeys]{revtex4-1} 
\usepackage{graphicx} 
\usepackage{dcolumn} 
\usepackage{bm} 
\usepackage{color} 
\usepackage{float} 
\usepackage{commath} 
\usepackage{amsmath} 
\usepackage{upgreek} 
\usepackage{physics} 
\usepackage[english]{babel} 
\usepackage[latin1]{inputenc} 
\usepackage{booktabs} 
\usepackage[unicode=true, bookmarks=true, bookmarksnumbered=false, bookmarksopen=false, breaklinks=false, pdfborder={0 0 1}, backref=false, colorlinks=false]{hyperref}
\usepackage[normalem]{ulem} 
\hypersetup{pdftitle={Phonon-polariton mediated dual electromagnetically induced transparency in a THz metamaterial}, pdfauthor={Amit Haldar, Kshitij V Goyal, Ruturaj Puranik, Vivek Dwij, Kanha Ram Khator, Subhashis Ghosh, Bhagwat S. Chouhan, Gagan Kumar, Satyaprasad P. Senanayak, Shriganesh Prabhu, and Shovon Pal}}

\begin{document}

\title{Phonon-polariton mediated dual electromagnetically induced transparency-like response in a THz metamaterial}

\author{Amit Haldar}
\altaffiliation{Both authors contributed equally}
\affiliation{School of Physical Sciences, National Institute of Science Education and Research, An OCC of Homi Bhaba National Institute (HBNI), Jatni, 752 050 Odisha, India}

\author{Kshitij V Goyal}
\altaffiliation{Both authors contributed equally}
\affiliation{School of Physical Sciences, National Institute of Science Education and Research, An OCC of Homi Bhaba National Institute (HBNI), Jatni, 752 050 Odisha, India}

\author{Ruturaj V Puranik}
\affiliation{Department of Condensed Matter Physics and Materials Science, Tata Institute of Fundamental Research, Mumbai, 400005 Maharashtra, India\looseness=-1}

\author{Vivek Dwij}
\affiliation{Department of Condensed Matter Physics and Materials Science, Tata Institute of Fundamental Research, Mumbai, 400005 Maharashtra, India\looseness=-1}

\author{Srijan Maity}
\affiliation{School of Physical Sciences, National Institute of Science Education and Research, An OCC of Homi Bhaba National Institute (HBNI), Jatni, 752 050 Odisha, India}

\author{Kanha Ram Khator}
\affiliation{School of Physical Sciences, National Institute of Science Education and Research, An OCC of Homi Bhaba National Institute (HBNI), Jatni, 752 050 Odisha, India}

\author{Subhashis Ghosh}
\affiliation{School of Physical Sciences, National Institute of Science Education and Research, An OCC of Homi Bhaba National Institute (HBNI), Jatni, 752 050 Odisha, India}

\author{Bhagwat S. Chouhan}
\affiliation{Department of Physics, Indian Institute of Technology Guwahati, Guwahati, 781 039 Assam, India}

\author{Gagan Kumar}
\affiliation{Department of Physics, Indian Institute of Technology Guwahati, Guwahati, 781 039 Assam, India}

\author{Satyaprasad P. Senanayak}
\affiliation{School of Physical Sciences, National Institute of Science Education and Research, An OCC of Homi Bhaba National Institute (HBNI), Jatni, 752 050 Odisha, India}

\author{Shriganesh Prabhu}
\affiliation{Department of Condensed Matter Physics and Materials Science, Tata Institute of Fundamental Research, Mumbai, 400005 Maharashtra, India\looseness=-1}

\author{Shovon Pal}
\altaffiliation{shovon.pal@niser.ac.in}
\affiliation{School of Physical Sciences, National Institute of Science Education and Research, An OCC of Homi Bhaba National Institute (HBNI), Jatni, 752 050 Odisha, India}

\date{\today}

\begin{abstract}
Light and matter can intertwine to create entirely new quantum states in the so-called strong-coupling regime, allowing unprecedented control over electromagnetic waves. In this work, strong-coupling mediated polaritonic states are exploited to demonstrate tunable dual electromagnetically induced transparency (EIT) like response in the terahertz (THz) frequency range using a micron-sized metamaterial system coupled with the phonon mode of a lead halide perovskite film. This architecture allows us to reversibly switch between the single and the dual EIT-like behavior without modifying the metamaterial structures. The dual EIT-like nature is further confirmed through the in-plane electric field distributions and the slow-light effects. The specific structural symmetry further allowed us to effectively switch between the dual EIT-like response and the conventional strong-coupling responses. Such tunability bears potential implications for developing metamaterial-phonon-based tunable THz devices such as switches, filters, and slow-light devices.
\end{abstract}

\maketitle


\section{Introduction}
The interaction of electromagnetic waves with matter is arguably one of the most important domains of modern physics, playing a pivotal role in the advancement of modern quantum technologies. Such interactions underpin the fundamental aspects of quantum optics and electrodynamics ranging from spontaneous emission to lasing\cite{Baranov2018ACSPhotonics}. The strength of this interaction is characterized by the coherent  exchange of energy between the light (in the form of a cavity) and the matter, reflected in the form of vacuum Rabi frequency $\Omega_{\rm R}$. The regime of so-called strong light-matter coupling is thus defined as the situation where the energy exchange between the light and the matter is much stronger than the rate at which the system tends to decohere. Effectively, the exchange rate should surpass the overall losses of the individual systems~\cite{Dovzhenko2018Nanoscale,Garcia2021Science,Torma2014RepProgPhys}. Under specific circumstances, the photonic (or the cavity) mode coherently hybridizes with material excitations, which could be excitons~\cite{Deng2010RMP}, phonons~\cite{Kim2020NanoLett,Zhang2021PRR} or even surface plasmons~\cite{Berini2012NatPho,Maier2006PRL}, thereafter creating new quasiparticles called the polaritons. These quasiparticles constitute the new hybrid light-matter eigenstates of the coupled system, displaying the anticrossing-like dispersion branches (namely, the lower and the upper polaritons) that are separated by the Rabi energy at the original point of crossing. Over the years, the hybrid light-matter states in the polaritonic systems have become very attractive, owing to their unparalleled potentials for quantum technologies, such as fast switching capabilities, strong nonlinear responses, and low power requirements for executing logic operations~\cite{Ballarini2013NatComm,Demirchyan2014PRL}. The manipulation of the hybrid quantum states, including their initialization, operation, and readout, have been achieved through pulsed-optical excitation at various energies~\cite{Bonadeo1998Science,Askarani2021PRL}. Incorporating polaritons into a tunable environment further introduces additional degrees of freedom, allowing deterministic control via electric~\cite{Lee2014AOM} and magnetic~\cite{Scalari2012Science} fields of the light, or even the mechanical perturbations~\cite{Vanner2015AP,Kim2023JPCL}. 

\begin{figure}[t!]
    \centering
    \includegraphics[width=\linewidth]{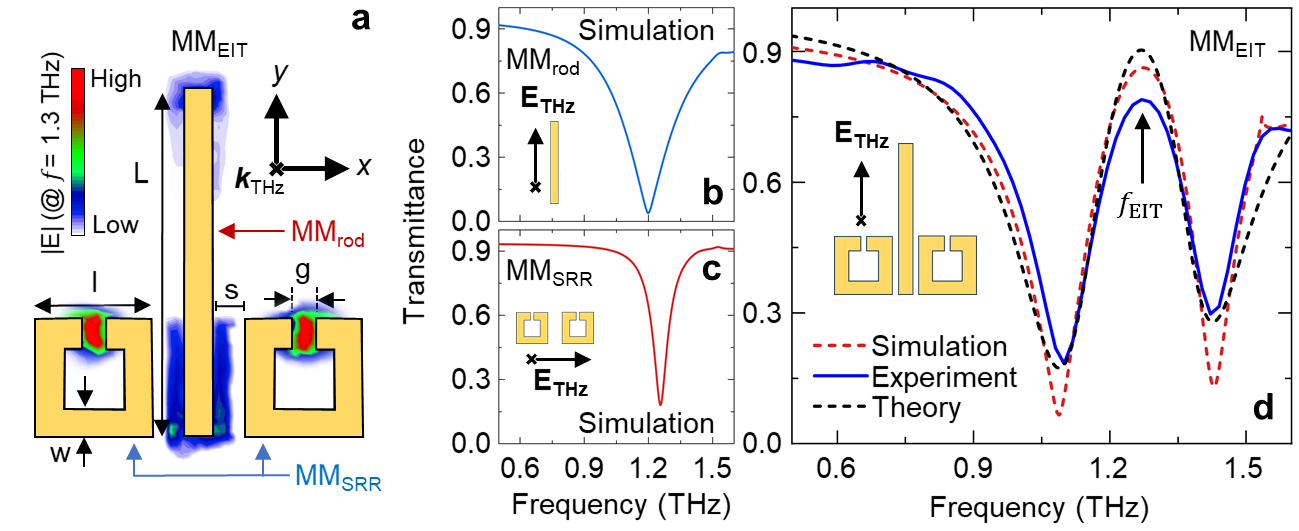}
    \caption{{\bf (a)} Schematic representation of the EIT-like metamaterial unit cell that consists of a rod resonator with a pair of split-ring resonator (SRR) on either sides of the rod. Geometric parameters of the metamaterial are: $\rm L\,=\,75\,\mu m$ , $\rm l\,=\,23.5\,\mu m$ , $\rm w\,=\,5\,\mu m$, $\rm g\,=\,3\,\mu m$, $\rm s\,=\,3\,\mu m$ with a periodicity along the $x$ and the $y$ directions as $\rm{\,67\,\mu m}$ and $\rm{\,93\,\mu m}$, respectively. The regions of in-plane field distribution, $|{\rm E}| = \sqrt{|{\bf E}_x|^2+|{\bf E}_y|^2}$ corresponds to the EIT peak frequency at 1.3\,THz. The magnitude of the electric field is indicated with the color map where red and white represents the high and low-field confinements, respectively. Simulated transmittance spectra of {\bf (b)} the rod and {\bf (c)} the SRR that shows the corresponding resonance frequencies at 1.2\,THz and 1.25\,THz, respectively. {\bf (d)} Simulated (red-dashed curve), measured (blue-solid curve) and theoretically obtained (black-dashed curve) transmittance spectra of the EIT-like metamaterial that shows the EIT peak frequency at $f_{\rm EIT}=1.3$\,THz, in an excellent agreement with each other. The insets in (b-d) show the incident electric-field polarization corresponding to the metamaterial structure.}
    \label{fig1}
\end{figure}

Artificially-engineered structures namely the metamaterials are evolving as promising alternatives to create a tunable environment for polaritonic systems. Metamaterials consist of a regular array of sub-wavelength planar metallic structures, which interact with the incident electromagnetic wave at the designed resonant frequency~\cite{Pendry1999IEEE,Pendry2006Science}. They have the capability to manipulate electromagnetic radiation in a way that allows realization of exotic properties, not present in natural materials such as negative refractive index~\cite{Shelby2001Science,Khan2024Optik} and sub-wavelength focusing~\cite{Kim2016SciRep}. Metamaterials act as a planar cavity and exhibit strong  localization and enhancement of electric fields around specific regions of the structure as per the design and the orientation of the electric field of the incident electromagnetic wave, as indicated in the schematic of Figure~\ref{fig1}a. The strong enhancement entails the effective coupling and hence the formation of hybrid polaritonic states~\cite{Kim2020NanoLett,Roh2023NanoLett}. 

One of the most celebrated phenomenon where the metamaterials have provided room-temperature classical analogues, is the electromagnetically induced transparency (EIT), which, however, is a quantum phenomenon. The EIT-like behavior results from the destructive interference between two seemingly different excitation pathways that concomitantly makes an initially opaque medium transparent to a probe laser beam, i.e., it generates a sharp transparency window within an absorption spectrum~\cite{Alzar2002AJP,Harris1997PT}. It is well known that one can achieve the EIT-like behavior in metamaterials primarily via two approaches. First and also the most relevant one for our work is the bright-dark mode coupling. Here, we couple the incident electromagnetic wave to the bright-mode resonator, characterized by a low Q-factor with high radiative nature. In contrast the incident wave is unable to excite the high Q-factor, dark-mode resonator. The dark mode, however, gets activated by the bright mode resonator via the near-field coupling~\cite{Singh2009PRB,Singh2014APL,Gu2012NC,Rao2017JIMTW,Manjappa2017OL,Devi2017OE,Yahiaoui2017APL}, as indicated in the schematic of Figure~\ref{fig1}a. For achieving the EIT-like behavior, the necessary condition is to couple the bright and quasi-dark resonances with different radiation rates by an appropriate engineering of the near-field coupling strength and resonator detunings with appropriate line-widths, leading to a destructive interference between the two modes. The other approach is the bright-bright mode coupling, where the two detuned bright modes are placed in close proximity for them to hybridize to result in an EIT-like peak~\cite{Yahiaoui2018PRB,Zhang2017OC,Vaswani2024OLT,Bhattacharya2021SciRep}. The EIT effect in this case is due to constructive/destructive interference~\cite{Fleischhauer2005RMP,Chiam2009PRB} leading to the appearance of maximum in transmission and suppression of reflection at EIT frequency. In the virtue of Babinet's principle same type of mechanism holds for complementary metasurfaces~\cite{Guo2012OE,Ortiz2021APL}, but this time with a maximum in reflection and minimum in transmission. The contrast of EIT effect is by far more pronounced since the absorption losses become negligible. Here, we categorically refer to our case as EIT-like. Moreover, while the bright-dark and bright-bright mode coupling frameworks are commonly used to explain the EIT-like behaviour, an alternative mathematical approach exists in the form of temporal coupled-mode theory (t-CMT)~\cite{Fan2003JOSA,Suh2004IEEE,Kodigala2015JAP}. This framework explicitly accounts for non-orthogonal modes, radiative decay rates, and the presence of exceptional points (EPs)~\cite{Liang2024OptExp,Kodigala2016PRB,Lawrence2014PRL}. We further note that the t-CMT model is typically applicable in the weak light-matter coupling regime, limiting its applicability for strongly coupled systems~\cite{Smith2025arxiv}.

The use of metamaterials for EIT has led to the development of new and ingenious applications in the THz frequency range, namely THz sensing~\cite{Xu2017Nanoscale} and switching~\cite{Hu2022ACSPhotonics}, THz modulators~\cite{Degl2022Nanophotonics}, and slow-light meta-devices~\cite{Manjappa2015APL}, operable at room temperatures. Tunable or dynamic EIT-like behavior is more captivating because of its potential in device applications. Tunability of metamaterials can be achieved in two different ways. In the first approach, one can introduce an auxiliary layer of active material such as  graphene~\cite{Kim2018ACSPhotonics,Liu2015NatComm}, vanadium dioxide~\cite{Wang2023APL}, or superconductors (for example NbN)~\cite{Li2021APL}. The conductivity and hence the carrier density of this layer can be easily tuned electrically using external voltage~\cite{Kim2018ACSPhotonics,Wang2023APL, Li2021APL} or even optically by using pulsed excitations~\cite{Hu2022ACSPhotonics}. Alternatively, one can also tune the metamaterials directly by changing the geometry (i.e., the structure)~\cite{Singh2009PRB,Chen2020OptMat} of the metamaterials or its orientation~\cite{Chiam2009PRB,Cheng2020AIPAdv,Sun2022OptExpress} with respect to the polarization of the incident electromagnetic radiation. The use of metamaterials has lately become a platform for generating phonon-polaritons both in the mid-infrared~\cite{Shelton2011Nanolett,Pons2019NatComm} as well as in the THz frequency range~\cite{Kim2020NanoLett,Kim2023JPCL}. With the growing demonstration of phonon-polaritons in the THz frequency range, the interplay between an EIT-like metamaterial and a phonon-polariton system remains to be explored, specifically if we can achieve any tunability of the said metamaterial system.

In this article, we exploit a combination of experimental and numerical routes to demonstrate the tunability from a single to a dual EIT-like behavior without changing the metamaterial structure in the THz frequency range. This is achieved by establishing a strong coupling between the THz EIT-like metamaterial cavity to a phonon mode. The phonon mode originates from a carefully selected lead halide perovskite, which maintains high crystallinity despite being deposited via a solution-based process. This approach enables easy integration with metamaterial structures and allows comparatively low-temperature processing while preserving a strong phonon response. In particular, we observe a transparency broadening in the coupled system. For different input polarization, the structural asymmetry of the metamaterial allows us to alter between the dual EIT-like response and the conventional strong-coupling response. In particular, we aim to implement a system with simultaneous slow-light and sub-wavelength confinement effects by integration of strongly coupled phonon with an EIT-like metamaterial, which has not been reported previously.

\section{Results and Discussion}
To study the phonon-polariton coupling in an EIT-like system, we start with a dimensionally modified version of a well-studied metamaterial structure~\cite{Liu2012APL}, as shown in Figure~\ref{fig1}a. The metamaterial unit cell consists of a pair of split-ring resonators (SRR) symmetrically placed on either side of the rod. To achieve the bright-dark mode EIT-like response, the rod and the SRR need to be in the near field coupling regime with appreciable resonance overlap~\cite{Liu2012APL,Chiam2009PRB,Liu2010NanoLett,Zhang2008PRL,Singh2009PRB}. The rod exhibits a dipole resonance ($f_{\rm rod}$) when directly excited by the incident THz pulse, polarized along its length (i.e., $y$-polarization) and hence acts as the bright mode, see Figure~\ref{fig1}b. The structural symmetry of the SRR, however, restricts its excitation by the incident $y$-polarized THz electric field. Instead, the SRR shows the LC-type resonance ($f_{\rm SRR}$) when the THz electric field is along the $x$-direction thereby acting as the dark mode (sub-radiant), shown in Figure~\ref{fig1}c. In the absence of the $x$-polarized incident THz electric field, the dark mode SRR pair is excited by the $x$-component of the THz electric field induced by the rod near its resonance. The field due to the dipole excitation of the rod and the capacitance gap of the SRR destructively interfere, thereby deactivating the resonance excitation of the rod. Such a deactivation results in an opening of a transparency window, leading to the EIT-like response, see Figure~\ref{fig1}d. The EIT-like metamaterial structure is fabricated on a 500\,$\mu$m thick $z$-cut quartz substrate of relative permittivity 4.41. Using standard maskless photo-lithography followed by thermal vapour deposition, a 120\,nm layer of aluminum was deposited. The fabricated metamaterial is then studied using terahertz time-domain spectroscopy (THz-TDS)~\cite{Pal2015SR,Mondal2024AOM,Puthukkudi2024AM}, with all measurements performed at room temperature. In addition, the simulations were performed using CST Microwave Studio and a two-coupled oscillator model was employed to get the theoretical plots~\cite{Chen2020OptMat,Ling2018Nanoscale,Cheng2013APL} for the EIT case(discussed in Supplementary Information section S8). Figure~\ref{fig1}d shows the simulated (red-dashed), measured (blue-solid) and the theoretically obtained (black-dashed) transmittance of the fabricated EIT structures, where a distinct transparency window, centered at the peak frequency $f_{\rm EIT} = 1.3$\,THz is observed. The transparency characteristics of the EIT obtained are comparable to other studies in literature exploring metamaterial-based EIT~\cite{Zhu2022JP}, with a transmittance of approximately 80\%. 

\begin{figure}[t!]
    \centering
    \includegraphics[width=1\linewidth]{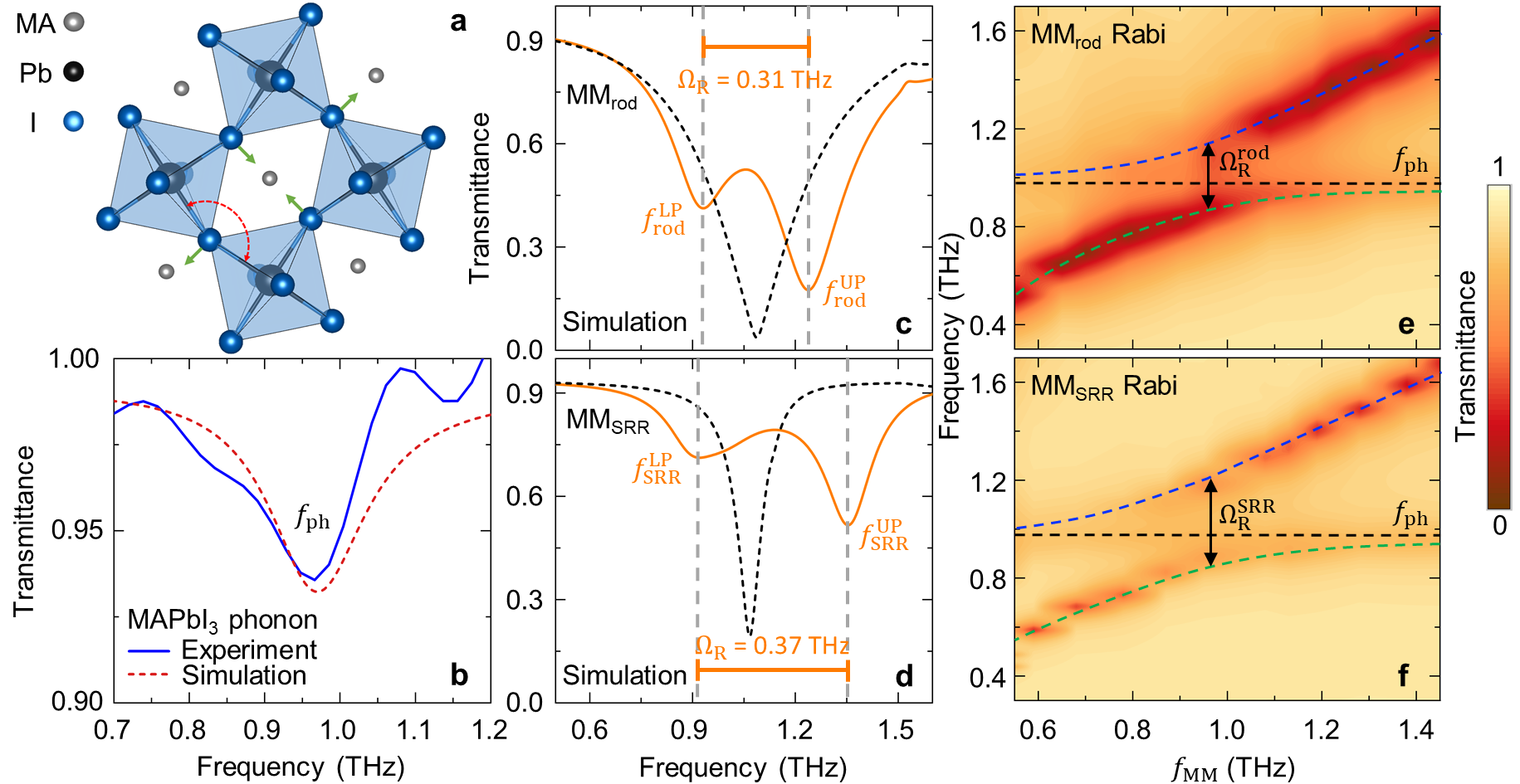}
    \caption{{\bf (a)} Schematic representation of MAPbI$_3$ perovskite structure with the arrows denoting the Pb--I--Pb angular bending corresponding to the transverse optical (TO) phonon mode. {\bf (b)} Simulated (red-dashed curve) and measured (blue-solid curve) transmittance spectra of the perovskite film spin-coated on a $z$-cut quartz substrate showing the phonon mode at $f_{\rm ph}=0.97$\,THz. Simulated transmittance spectra of {\bf (c)} the rod and {\bf (d)} SRR in the presence of the MAPbI$_{3}$ layer with phonon mode, showing the formation of upper ($f^{\rm UP}_{\rm MM}$) and lower ($f^{\rm LP}_{\rm MM}$) polaritons. Here, MM corresponds to either the rod or the SRR. The black-dashed curves represent the corresponding resonator response in the absence of phonon mode. Separation between the polaritonic states is indicated by the corresponding Rabi splitting $\Omega_{\rm R}$. Simulated 2D contour plot of the transmittance obtained by varying resonance frequency of the {\bf (e)} rod and {\bf (f)} SRR metamaterial structures. The black-dashed lines denotes the MAPbI$_3$ phonon mode at 0.97\,THz. Blue- and green-dashed curves denote a guide-to-the-eye representation for the upper and lower polaritonic states, respectively. For the SRR structures the incident THz electric field polarization was kept along $x$ direction while that for the rod structures it was along the $y$ direction.}
    \label{fig2}
\end{figure}

To obtain a phonon-polariton system, we have used a lead-halide perovskite, namely methylammonium lead triiodide (MAPbI$_{3}$) of thickness 200\,nm, the structure of which is schematically shown in Figure~\ref{fig2}a. The film thickness is optimized to ensure a balance between the volume fraction of metamaterial near-field coupling with MAPbI$_{3}$ and the material transmittance. MAPbI$_{3}$ is known to exhibit a transverse optical (TO) phonon mode at around 1\,THz~\cite{Sendner2016MaterHoriz,LaOVorakiat2016JPCL}. This mode corresponds to the octahedral distortion involving a change in the Pb--I--Pb bond angle. We performed the THz-TDS characterization of MAPbI$_{3}$, spin-coated on the $z$-cut quartz substrate. To make the phonon mode in MAPbI$_3$ active, we then anneal~\cite{Leguy2016PCCP,Kim2017NatComm} the system to 100$^{\circ}$C for 15 minutes. The blue-solid curve in Figure~\ref{fig2}b shows the experimentally observed phonon mode at 0.97\,THz that closely matches the one obtained numerically by optimizing the relevant parameters in the CST-Microwave Studio (red-dashed curve).

The formation of phonon-polariton was numerically established using the individual elements of the EIT-like metamaterial. Figures~\ref{fig2}c and~\ref{fig2}d show the simulated transmittance response of the rod and SRR, when the active phonon mode of MAPbI$_3$ hybridizes with the corresponding metamaterial resonances. Note that with the introduction of the MAPbI$_3$ the effective refractive index changes and this red-shifts the original metamaterial resonance. For getting an estimate of this effect, we have simulated the MAPbI$_3$-metamaterial system, with the phonon of MAPbI$_3$ deactivated. The corresponding resonances are shown in black-dashed lines Figures~\ref{fig2}c and~\ref{fig2}d. Numerically, we activated the phonon mode by introducing a Lorentz oscillator in addition to the dielectric properties. Upon activation, we find that the phonon couples with the metamaterial mode leading to the formation of polaritons that are separated by Rabi frequency, which quantifies the coupling strength between the systems. For the rod, the lower and upper polaritons are found at 0.9\,THz and 1.2\,THz, respectively, with a Rabi frequency of 0.31\,THz. In contrast the SRR shows a slightly higher Rabi frequency of 0.37\,THz due to the stronger field confinement in the SRR gap, with the polaritonic states at 0.9\,THz and 1.3\,THz. Experimentally, a Rabi splitting value of 0.26\,THz is seen for the SRR structure which is comparable to phonon based strong coupling studies~\cite{Kim2020NanoLett,mekonen2024ACSPhot}.

By tuning the resonance frequency $f_{\rm MM}$ of the rod and the SRR, the avoided crossing behavior is studied to examine the emergence of the upper and the lower polaritons. Figures~\ref{fig2}e and~\ref{fig2}f show the simulated 2D contour plots of transmittance for the individual rod and SRR Rabi splitting. In both cases, the periodicity along $x$ and $y$ directions were systematically changed along with the length $L$ for the rod while the length $l$ for the SRR. The blue- and green-dashed curves in Figures~\ref{fig2}e and~\ref{fig2}f are a guide-to-the-eye indicators for the upper and the lower polaritonic states, respectively. The black dashed line denotes the MAPbI$_3$ phonon mode at 0.97 THz. It can be seen that a greater Rabi splitting is observed for the SRRs owing to the stronger field confinement SRR capacitive gap~\cite{Kim2020NanoLett,Roh2023NanoLett}. 

It is now imperative to first establish that the hybridized phonon-polariton is in the strong-coupling regime. Along that direction, we evaluate the interaction potential using the coupled oscillator model~\cite{Torma2014RepProgPhys,Liu2015NatPhotonics,Kim2020NanoLett}, where the interaction Hamiltonian is given by,
\begin{equation}
    \begin{bmatrix} 
        E_{\rm MM} + i\hbar\Gamma_{\rm MM} & V \\ 
        V & E_{\rm ph} + i\hbar\Gamma_{\rm ph}.
    \end{bmatrix}
\end{equation}
Here, $E_{\rm MM}=2\pi\hbar f_{\rm MM}$ is the resonant energy of the metamaterial. Note that MM refers to either rod or SRR. $V$ is the interaction potential and $E_{\rm ph}=2\pi\hbar f_{\rm ph}$ is the phonon energy. $\hbar\Gamma_{\rm MM}$ and $\hbar\Gamma_{\rm ph}$ are the half-width at half-maximum (HWHM) values of the metamaterial and the phonon resonances, respectively. Under the framework of this model, when $E_{\rm MM}=E_{\rm ph}$, the interaction potential $V$ is given by~\cite{Kim2020NanoLett}:
\begin{equation}
    V=\frac{1}{2}\sqrt{(\hbar\Omega_{\rm R})^2+(\hbar\Gamma_{\rm ph}-\hbar\Gamma_{\rm MM})^2}.
\end{equation}
Here, $\hbar\Omega_{\rm R}$ is the energy shift due to the Rabi splitting. To enter the strong coupling regime, $V>\sqrt{((\hbar\Gamma_{\rm ph})^2+(\hbar\Gamma_{\rm MM})^2)/2}$. Considering the phonon-SRR system, from the experimental results, we find that $\hbar\Gamma_{\rm ph}=0.323$\,meV and $\hbar\Gamma_{\rm SRR}=0.261$\,meV, as indicated in Figure~S4 of the Supplementary Information. This implies that the threshold for strong coupling ($V_{\rm st}$) in the SRR resonance is 0.294\,meV. The energy of the Rabi splitting is calculated from the energy shift seen in the transmittance spectra of the SRR-phonon system (see Figure~S4 of the Supplementary Information), which is 1.075\,meV, giving an interaction potential of 0.538\,meV~$>V_{\rm st}$. This clearly substantiates that experimentally we are in the strong coupling regime for SRR metamaterial, although the observed values for the Rabi splitting is lower in our experiments compared to our numerical simulations, which can be attributed to structural imperfections of the metamaterial and/or undulations in the deposited perovskite film. Similar calculations also validate the strong coupling regime for the rod metamaterial.

\begin{figure}[t!]
    \centering
    \includegraphics[width=0.9\linewidth]{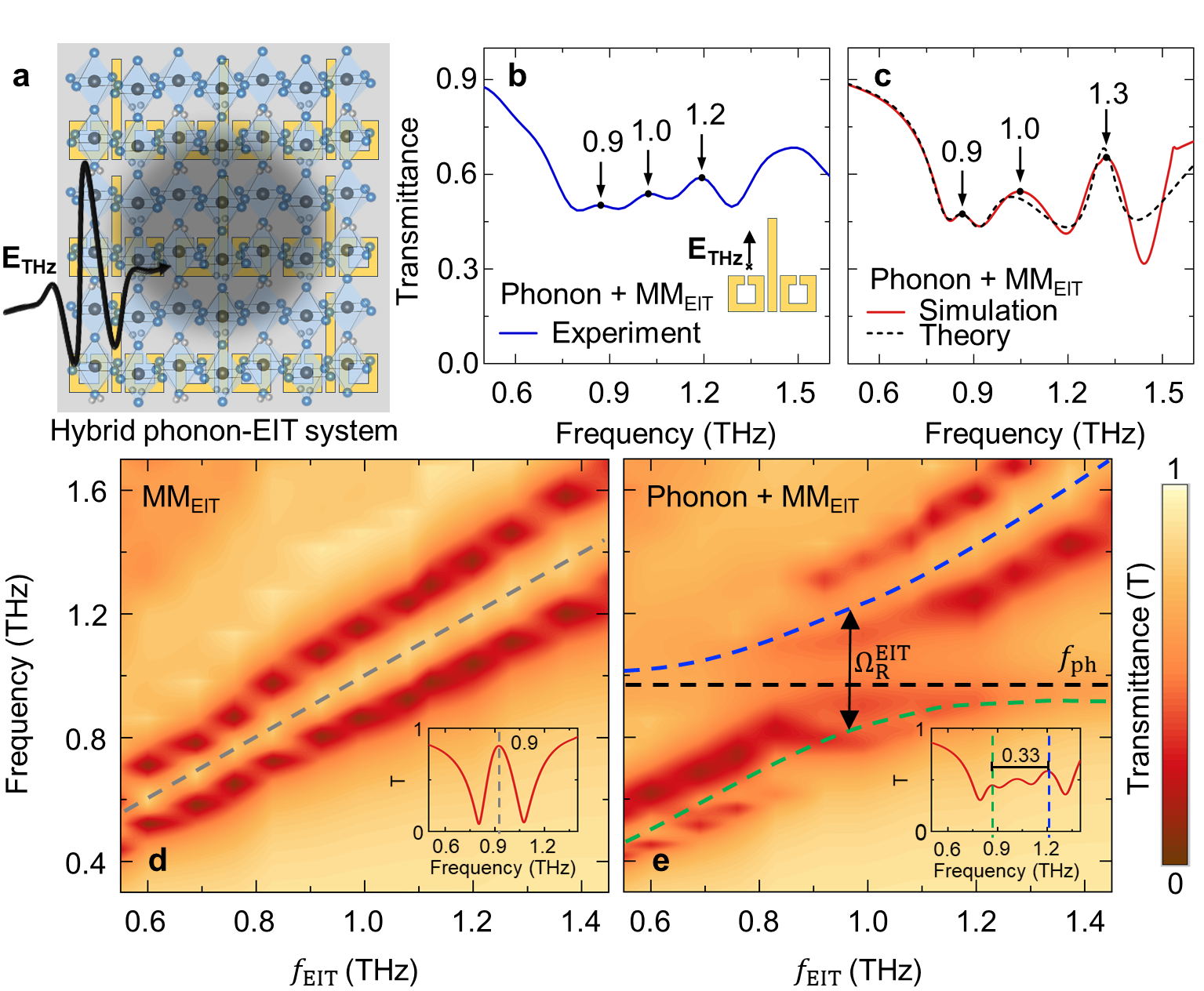}
    \caption{{\bf (a)} Schematic representation of the hybrid phonon-EIT metamaterial structure, where the phonon mode results from the spin-coated, crystallized MAPbI$_3$ perovskite layer. {\bf (b)} Measured and {\bf (c)} simulated (red-solid curve) and theoretically obtained (black-dashed curve) transmittance spectra of the hybrid phonon-EIT system in the strong coupling regime. The transmittance peaks at 0.9\,THz, 1.0\,THz, and 1.2\,THz are observed due to the splitting of the EIT-like response, induced by the phonon mode of MAPbI$_{3}$. Simulated 2D contour plots of the transmittance for the hybrid phonon-EIT structure when the phonon mode is {\bf (d)} deactivated and {\bf (e)} activated. The presence of phonon leads to an  avoided crossing feature signaling the concurrence of both strong coupling and EIT-like effects. While the gray-dashed line in (d) represents the EIT peak, the black-dashed line in (e) denotes the MAPbI$_3$ phonon mode. Blue- and green-dashed curves in (e) denote a guide-to-the-eye representation for the upper and lower polaritonic-EIT states, respectively. The insets in (d) and (e) show typical transmittance spectra along $f_{\rm{EIT}}= 0.9$\,THz.}
    \label{fig3}
\end{figure}

We now move on to the hybrid system that comprises of the phonon mode from MAPbI$_3$ and the EIT-like metamaterial, which we henceforth refer to as the hybrid phonon-EIT structure, see the schematic in Figure~\ref{fig3}a. It is to be noted that our original choice of structural parameters to result in an $f_{\rm EIT} = 1.3$\,THz was intentional. The choice was made to compensate for the red-shift induced by the high permittivity of the spin-coated MAPbI$_3$ layer~\cite{Kim2020NanoLett}, as discussed earlier, where the EIT peak frequency shifts from 1.3\,THz to 1.08\,THz, very close to the target phonon frequency of 0.97\,THz. Figures~\ref{fig3}b and~\ref{fig3}c show the experimentally observed, simulated (red-solid) and the theoretically obtained (black-dashed) THz transmittance spectra of the hybrid phonon-EIT system, respectively, which are in a good qualitative agreement. The theoretical plots are obtained using a modified version of three coupled oscillators, which is further discussed in the Supplementary Information, Section S8. A substantial increase in the width of the transparency window is seen (more evident in the inset of ~\ref{fig3}e), albeit at the cost of the overall transmittance amplitude due to the absorption in the perovskite layer. The trade-off between the bandwidth and the transmittance can be obtained either by further optimization of the MAPbI$_3$ layer thickness or improving the crystallinity of MAPbI$_3$. This would minimize the absorption losses and thereby improving the transmittance while maintaining sufficient phonon activity for strong coupling. In addition, choosing a material with a sharper phonon mode can considerably improve the peak amplitudes, as demonstrated hypothetically in Figure~S6 of the Supplementary Information. The single EIT peak splits into two distinct transparency windows with corresponding peaks at 0.9 and 1.2\,THz. The peak at 1\,THz in the transmittance curve, however, does not show EIT-like behavior, but a simple superposition of the centers of the two Rabi splitting, which we will substantiate later. To understand the emergence of the dual transparency peaks, an order of excitation caused by the input THz pulse has to be established. At first, the rod is directly excited by the incident $y$-polarized THz electric field. This leads to local field enhancement near its ends that couple with the phonon mode to give the Rabi splitting, with the corresponding upper ($f_{\rm rod}^{\rm UP}$) and lower ($f_{\rm rod}^{\rm LP}$) polaritonic states. Thereafter, the SRR is excited because of the resonances emerging from the rod's polaritonic states, effectively coupling to the strongly-coupled rod-phonon system. The excited SRR resonance then creates a strong local field within the SRR gap, substantially exciting the phonon mode and producing an additional Rabi splitting, with its corresponding upper ($f_{\rm SRR}^{\rm UP}$) and lower ($f_{\rm SRR}^{\rm LP}$) polaritonic states. The pair of emergent polaritonic states now resemble overlapping bright-dark EIT-like resonances. More precisely, $f_{\rm rod}^{\rm UP}$ (bright) and $f_{\rm SRR}^{\rm UP}$ (dark) overlap to form what we can refer to as the upper-polaritonic-EIT state at 1.2\,THz and in the same fashion we get the lower-polaritonic-EIT state at 0.9\,THz. A destructive interference of the net charge distribution is observed that gives rise to dual transparency window at the corresponding frequencies, as shown in Figures~\ref{fig3}b and~\ref{fig3}c. The higher frequency peak shows greater transparency because of the overlapping of higher transmission dips. 

Figure~\ref{fig3}d and~\ref{fig3}e show the simulated 2D contour plots of transmittance by varying the $f_{\rm EIT}$ in the absence and the presence of the TO phonon mode of MAPbI$_3$. These plots are simulated by varying the structure dimensions (more precisely, the periodicity in $x$ and $y$ directions along with the lengths L and l of the rod and SRR, respectively) to scan across the EIT peak frequency. A clear avoided-crossing-like behavior is observed that can be attributed to the splitting of a single $f_{\rm EIT}$ peak into upper-polaritonic- and lower-polaritonic-EIT peaks when the phonon mode is active in the system, evident from the blue- and the green-dashed curves in Figure~\ref{fig3}e. Equivalently, four distinct resonance dips are obtained, as seen in Figure~\ref{fig3}e, especially near the phonon resonance at 0.97\,THz. According to our interpretation of the polaritonic dips leading to the emergence of EIT-like peaks, the effective Rabi frequency in the combined system is simply the frequency difference between the two observed peaks. Interestingly, applying this definition to the combined system for $f_{\rm EIT}$ = 0.9\,THz, yields an effective increase in the splitting to 0.33\,THz compared to the individual rod and SRR case, shown in the inset of Figure~\ref{fig3}e. Such an increase can be attributed to the slow light effect induced due to the EIT-like behavior, thereby increasing the net splitting for the slowed-down cavity mode frequency. Notably, by deactivating or activating the phonon mode, we can switch from a single to a dual EIT-like nature -- a feature that has been reported recently by rotating the metamaterial structure~\cite{Hu2022ACSPhotonics}, allowing controlled manipulation of THz wavefronts.

\begin{figure*}[t!]
    \centering
    \includegraphics[width=\linewidth]{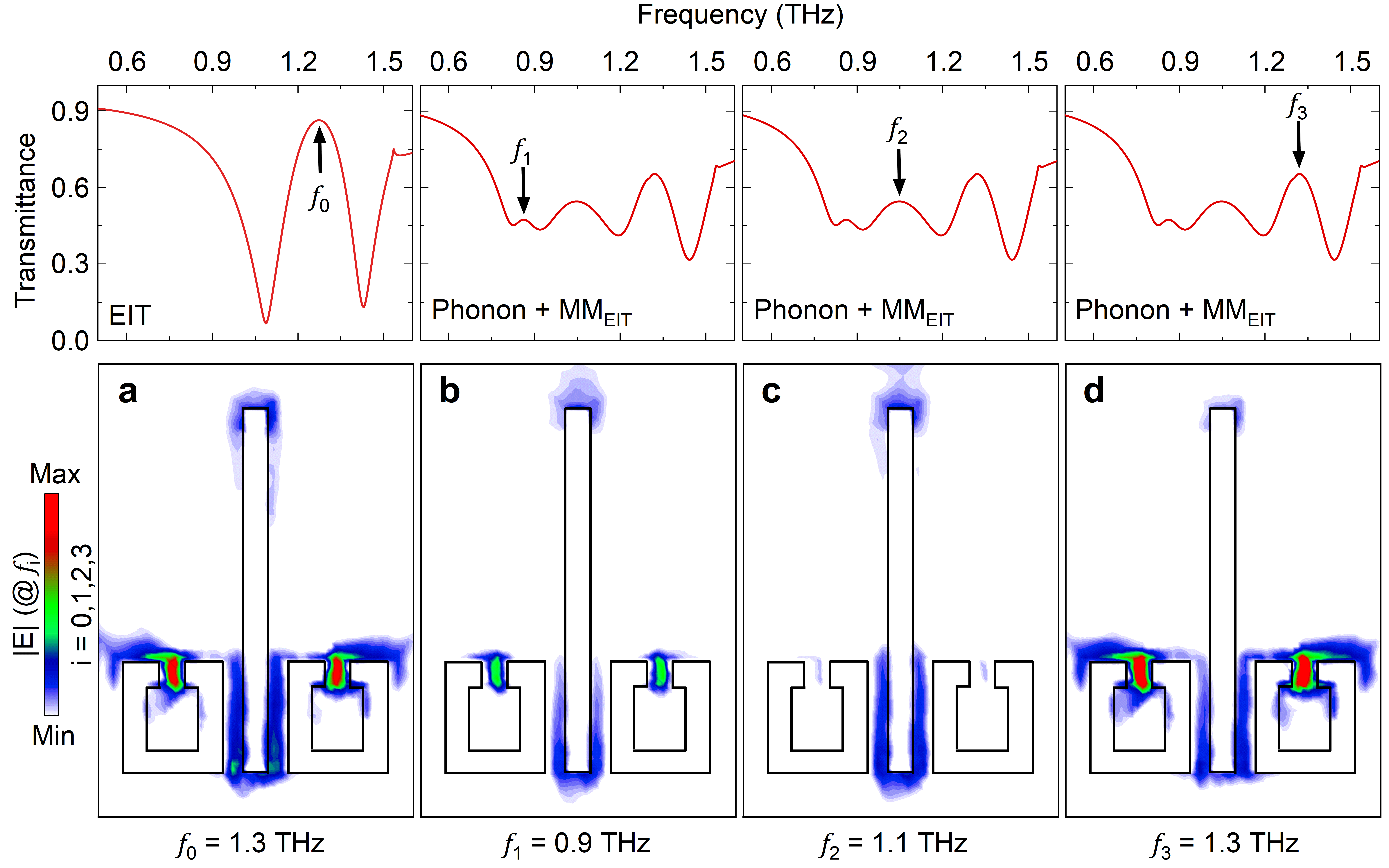}
    \caption{Normalized in-plane electric field distribution, $|{\rm E}| = \sqrt{|{\bf E}_x|^2+|{\bf E}_y|^2}$ corresponding to $f_i$ for {\bf (a)} EIT only structure without MAPbI$_3$, i.e., at $f_0=1.3$\,THz and {\bf (b-d)} with the strong coupling in the EIT-like system upon the activation of phonon mode in MAPbI$_3$ at $f_1=0.9$\,THz, $f_2=1.1$\,THz, and $f_3=1.3$\,THz, respectively. The normalization is done with respect to the maximum in-plane field amplitude for the EIT only structure at  $f_0=1.3$\,THz.}
    \label{fig4}
\end{figure*} 

As discussed earlier, the bright-dark mode EIT-like effect is achieved by deactivating the absorption of the rod due to the destructive interference between the oppositely directed electric fields produced by the rod's dipole and SRR gap at the resonance. We use the normalized in-plane electric field distribution across the metamaterial structures to validate the EIT-like nature of the peaks at 0.9\,THz and 1.3\,THz, corresponding to the lower-polaritonic- and upper-polaritonic-EIT peaks. Note that for the experimental situation, the upper-polaritonic-EIT peak is at 1.2\,THz, see Figure~\ref{fig3}b. Figure~\ref{fig4}a clearly shows the distinct field distribution corresponding to the 1.3\,THz EIT-like peak in the absence of the phonon activation. The dipole resonance of the rod is deactivated, preventing the resonant absorption of the $y$-polarized input THz field and the in-plane field is primarily located in the SRR gap. A similar deactivation is also observed for the peaks obtained at 0.9\,THz (Figure~\ref{fig4}b) and 1.3\,THz (Figure~\ref{fig4}d), suggesting the EIT-like effect. The nature of the in-plane field in Figure~\ref{fig4}b also suggests the lower magnitude of the opposing fields in the rod and SRR, indicating the partial deactivation of the rod and hence explaining the reduction in the EIT-like peak amplitude. At 1.1\,THz, however, the opposing field in the SRR is absent, which implies the lack of EIT-like nature and rather suggest a superposition of the Rabi splitting between the two components of the EIT structure, namely the rod and the SRR.

\begin{figure*}[t!]
    \centering
    \includegraphics[width=0.75\linewidth]{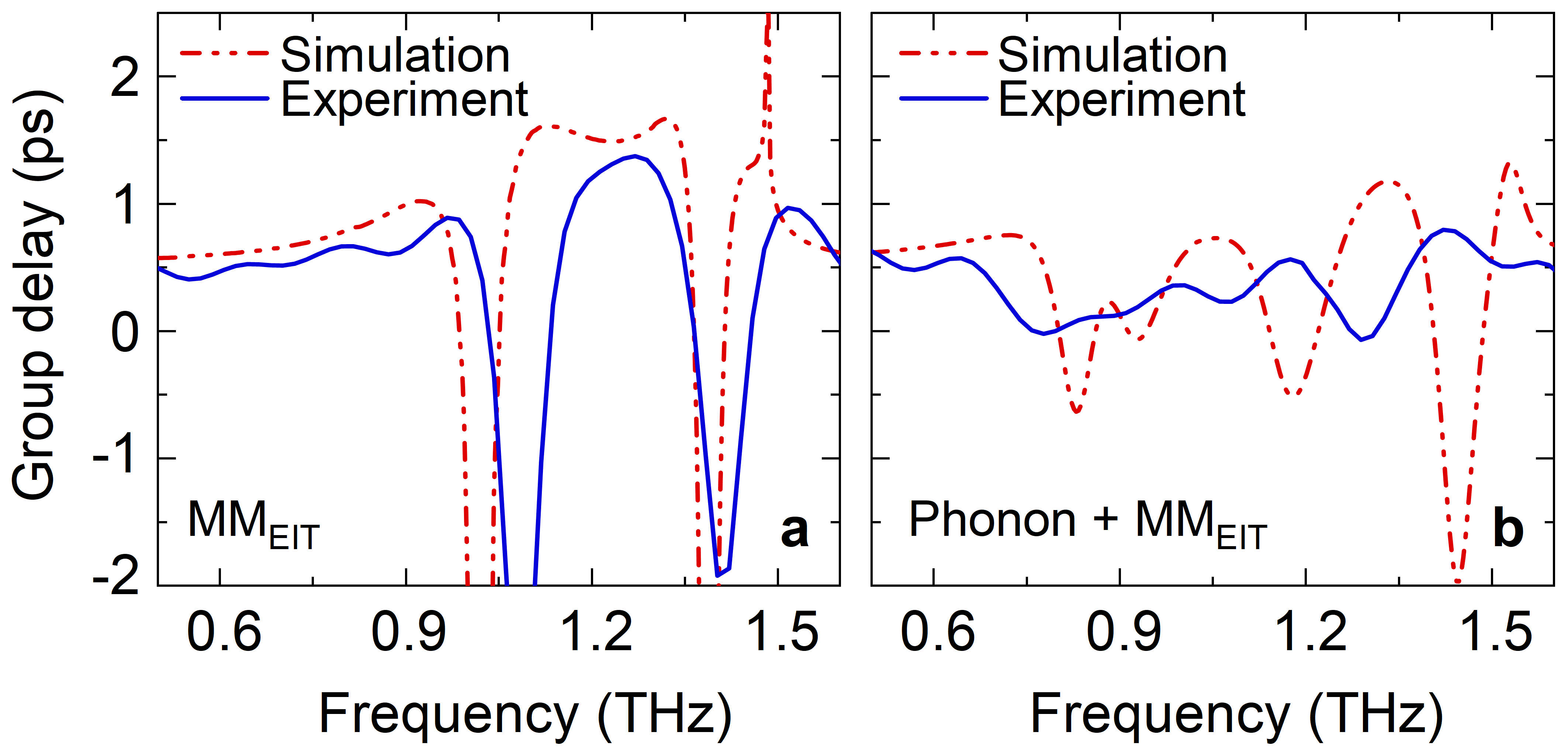}
    \caption{Simulated (red-dashed curve) and experimentally observed (blue-solid curve) group delay plots for {\bf (a)} EIT only structure without MAPbI$_3$, and {\bf (b)} with strong coupling in the EIT-like system upon the activation of the phonon mode in MAPbI$_3$. A delay of approximately 1.5\,ps is observed in the simulation for EIT-like metamaterial at the EIT peak frequency. Introduction of phonon mode causes additional delay effects at 0.9\,THz and 1.3\,THz, otherwise absent in the absence of phonons.}
    \label{fig5}
\end{figure*}

To further scrutinize the EIT-like effect, we show the frequency variation of the group delay before and after the deposition of phonon-active MAPbI$_3$. EIT-like metamaterials, display a characteristic group delay owing to the narrow absorption dips that are associated with a rapid change in the dispersion relation and hence the refractive index, as seen from the Kramers-Kronig relations~\cite{Kash1999PRL,Hau1999Nature,Tamayama2012PRB}. The detection in THz-TDS being phase sensitive, we were able to map the phase changes ($\phi$) and consequently calculate the group delay $\tau_{\rm g}$ using the relation~\cite{Sarin2023ResultsOpt, Woodley2004PRE,Hu2022ACSPhotonics,SunOptExp2022},
\begin{equation}
    \tau_{\rm g}=-\frac{1}{2\pi}\frac{d\phi}{df}
\end{equation}
The slow light effect for the EIT-like metamaterials is clearly evident from Figure~\ref{fig5}a, where we achieve a group delay of about 1.5\,ps at around 1.3\,THz, with a good qualitative agreement between our experiments and simulations. A group delay of 1.5\,ps is not particularly high with respect to values found in literature~\cite{Zhu2022JP,Gu2012NC} for metamaterial-based EIT structures but is sufficient to demonstrate the slow-light effect in the hybrid system. Introducing the phonon mode shows a group delay that undergoes splitting (i.e., a single group delay peak now splits into three), which is expected from a dual EIT-like response~\cite{Hu2022ACSPhotonics,Li2020Mat,Zhu2020ApplPhysA}, shown in Figure~\ref{fig5}b. Initially, no group delay was present at 0.9\,THz and 1.3\,THz but in the combined system, group delay peaks emerge at frequencies where the EIT-like nature is predicted to occur. Additionally, the negative group delay response visible in the dips in Figure~\ref{fig5}b adds another layer of pulse shaping effects in the system exhibiting anomalous dispersion~\cite{Woodley2004PRE,Dolling2006Sci,Sarin2023ResultsOpt}. Such a system, as we speculate, can have direct applications in custom multi-channel slow-light devices, tunable beam shaping surfaces and switchable delay response materials via rotation or translation to the metamaterial or even by the modulation of phonon. Compared to other dual-EIT devices~\cite{Hu2022ACSPhotonics,Zhu2021OME}, our system exhibits a relatively lower EIT amplitude, primarily due to the incorporation of a much weaker phonon resonance as opposed to the stronger metamaterial resonances employed in those studies. However, our approach uniquely integrates two fundamentally distinct physical mechanisms to achieve the dual-transparency window, rendering direct performance comparisons less meaningful.

\begin{figure*}[t!]
    \centering
    \includegraphics[width=0.9\linewidth]{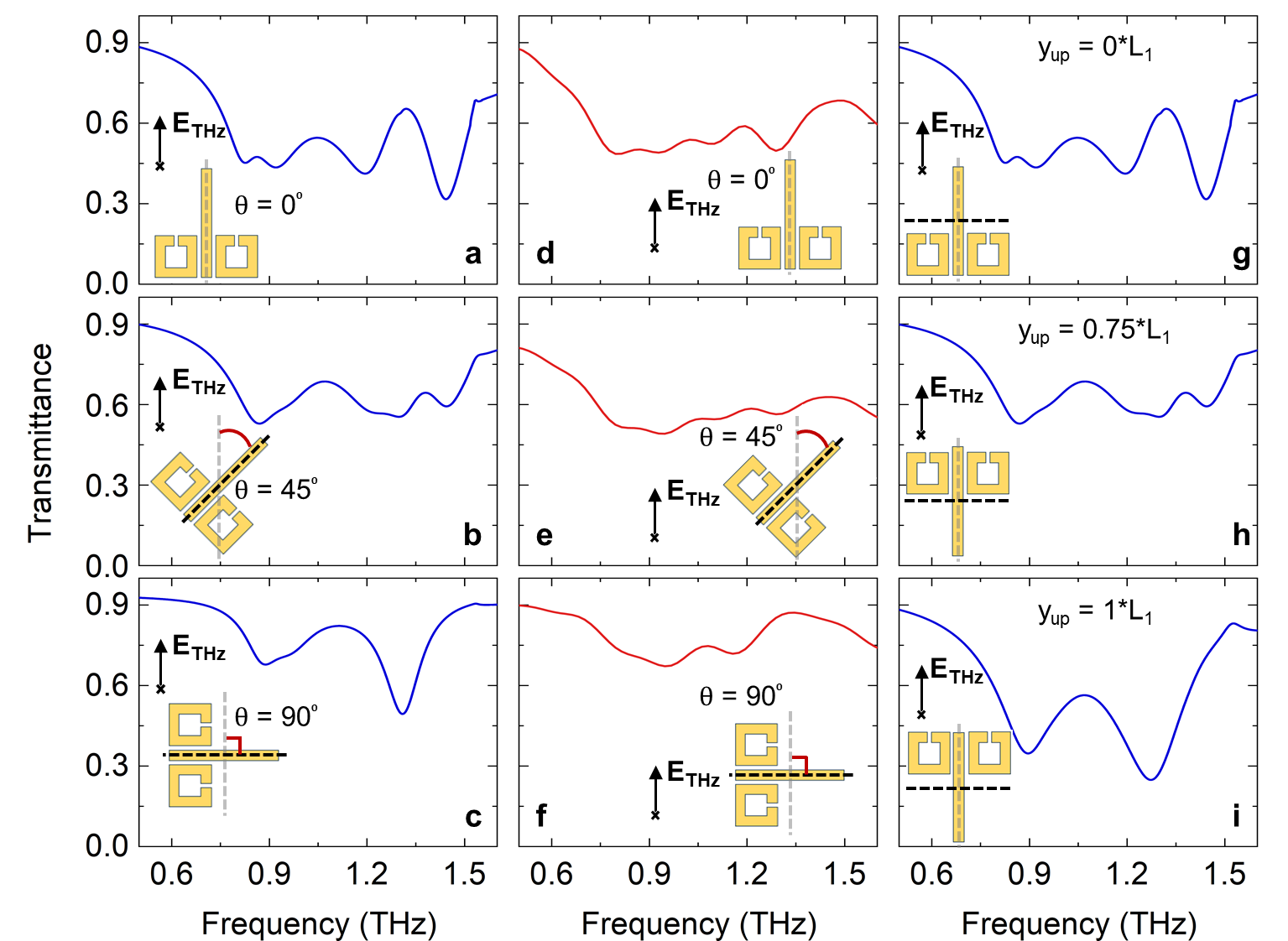}
    \caption{Simulated {\bf (a-c)} and experimental {\bf (d-f)} transmittance spectra of the phonon-EIT hybrid metamaterial structure with  rotational tunability for $\rm{\theta}=0^{\circ},45^{\circ}$, and $90^{\circ}$. Simulated spectra of phonon-EIT hybrid metamaterial structure {\bf (g-i)} for translational tunability. $y_{\rm up}$ is the magnitude of the upward translation of the SRR along the length of the rod in $\mu$m. Here the reference point for translation is the center of the SRR at $y_{\rm up}=0$. We define $\rm{L_1}=L-l$.}
    \label{fig6}
\end{figure*}

We have further explored the rotational (both numerically and experimentally) as well as the translational (numerically) tunability that allows us to switch the response of the hybrid system from the combined effect to a traditional Rabi splitting. Such tunability in the EIT-only metamaterial has been demonstrated previously~\cite{Liu2012APL}. By rotating the polarization of the input light or equivalently rotating the metamaterial, the bright mode of the rod can be effectively deactivated (see Figure~S7a-c of the Supplementary Information), leaving us with the Rabi-split polaritons of the SRR structure at 90$^{\circ}$ rotation. Figures~\ref{fig6}a-c and~\ref{fig6}d-f, represent the simulated and the experimental results for the hybrid phonon-EIT system, where the structure is rotated. The disappearance of EIT-like effect in the system facilitates a dynamic transition from the phonon-polariton-induced dual-EIT-like nature to a conventional phonon-polaritonic system (phonon-MM$_{\rm SRR}$). Alternately, the entire interaction can be viewed as a hybridization between three coupled oscillators, and by rotating the sample, we deactivate one of the oscillator. The continuous nature of this tunability without any structural modifications gives us a map of the changes in a phonon-polariton system as the EIT-like effect sets in. Specifically, the rotational tunability in our design provides access to a regime where one can switch between two fundamentally different optical effects i.e. strong coupling and dual-EIT effect, simply by rotating the incident polarization. This is not just a binary toggle but a continuous control mechanism. Such control acts as a finely tunable optical "knob", allowing for systematic modulation of the resonator-resonator interaction strength.

In a similar fashion, the upward translation of the SRR pair prevents the formation of EIT-like nature~\cite{Liu2012APL} (see Figure~S7d-f in the Supplementary Information). This is because of the destructive interference of the electric and the magnetic pathways. For the hybrid structure, this implies that the translation switches towards the phonon-polariton Rabi splitting of the rod structure, the results of which are shown in Figure~\ref{fig6}g-i. The upward translation value determines the percentage of deactivation of the SRR's dark mode. At $y_{up}=L_1$, the resonance vanishes and the phonon-polaritonic response of the rod is observed, see Figure~\ref{fig6}i. Translational tunability, however, involves alteration of the metamaterial structure.

While the rotational tunability enables the transition from the combined effect in the hybrid system to the strong coupling effect in SRR, by translation, we switch between the combined effect and strongly coupled rod-phonon system. Both these methods provide feasible mechanisms to modulate between the phonon-polariton-induced dual-EIT and conventional phonon-polaritons of the individual rod or SRR metamaterial.

\section{Conclusions and Outlook}
In conclusion, we have presented a comprehensive study on a hybrid phonon-EIT metamaterial system where we have demonstrated the phonon-polariton mediated dual EIT-like response. Such a device can not only harnesses the slow-light effect but also confine sub-wavelength cavity modes of the electromagnetic fields, offering enhanced control of light-matter interaction at the micro-scale. The transition from a single EIT-like peak to a dual EIT-like peak is achieved without any alteration of the metamaterial structure itself. The dual EIT, in contrast, is solely governed by the strong coupling effects, mediated by the phonon mode of MAPbI$_{3}$. The dual EIT-like response led to a substantial increase of the effective transparency window, however, at the cost of the net transmittance. We have numerically substantiated the EIT-like nature of the peaks via the in-plane electric field distribution across the metamaterial structure. Taking the advantage of the structural asymmetry we have reversibly switched between the dual EIT-like and the conventional strong coupling regimes by rotating the hybrid structure with respect to the polarization of the incident light. Essentially, this is much like a control knob, allowing us to vary the strength of the EIT-like effect in the hybrid system and thereby change the overall interplay of the coupling, paving the way for a multifunctional device on a single platform. For example, the dual EIT response demonstrated in our system enables selective transmission of THz frequencies, which can be dynamically tuned via rotational or translational adjustments or by modulating the phonon mode. This can be exploited to fabricate switchable narrow-band THz filters or frequency-selective switches without needing structural reconfiguration, crucial for flexible THz communication and sensing technologies. Similarly, the appearance of two narrow transparency windows allows simultaneous detection of multiple spectral features, enabling multi-channel THz sensing~\cite{Zhu2021OME} for chemical and biological detection applications. In addition, the strong group delay observed at the transparency windows (up to ~1.5 ps) suggests that our hybrid structure can act as a THz slow-light device. Such devices are important for buffering THz signals in real-time data processing, pulse shaping, and delay lines. The simple device architecture involving maskless photo-lithography of meta-structures and  solution-based spin-coating of halide perovskite further allows large-scale fabrication and device integration. Our results thus demonstrate a new avenue to explore the reversible tunability in EIT-like metamaterials without incorporating any irreversible structural modifications. Phonon-polariton systems have been known to enhance nonlinear susceptibility~\cite{Suresh2020ACSPhotonics} and induce second harmonic generation~\cite{Razdolski2018PRB,Razdolski2016NanoLett}. An integration with the EIT-like effect can bring about a spectral selectivity and increase lifetimes of such light-matter interactions.

\vspace{5pt}
{\noindent \bf SUPPORTING INFORMATION}\\
The supporting information contains details on terahertz time-domain spectroscopy (THz-TDS) setup, fabrication of metamaterial, deposition of MAPbI$_3$, EIT tunability, and information on the energy-level diagram for the hybrid system.

\vspace{5pt}
{\noindent \bf ACKNOWLEDGEMENT}\\
A.H., K.V.G., K.R.K, S.G., S.P.S, and S.P. acknowledge the support from DAE through the project Basic Research in Physical and Multi-disciplinary Sciences via RIN4001. S.P. also acknowledges the start-up support from DAE through NISER and SERB through SERB-SRG via project no.~SRG/2022/000290. S.M. acknowledges INSPIRE-SHE for the financial support. S.P.S. acknowledges the funding from Royal Society London, UK and ANRF-IRHPA funding via project no.~ANRF/IPA/2021/000096. S.S.P. acknowledges the funding support from DAE through the project RTI4003.

\vspace{5pt}
{\noindent \bf COMPETING INTERESTS}\\
The authors declare that they have no competing financial interests.

\vspace{5pt}
{\noindent \bf DATA AVAILABILITY}\\
The experimental and numerical datasets are available from the corresponding author upon reasonable request.

\end{document}


\title{Supplementary Information: Phonon-polariton mediated dual electromagnetically induced transparency-like response in a THz metamaterial}

\author{Amit Haldar}
\altaffiliation{Both authors contributed equally}
\affiliation{School of Physical Sciences, National Institute of Science Education and Research, An OCC of Homi Bhaba National Institute (HBNI), Jatni, 752 050 Odisha, India}

\author{Kshitij V Goyal}
\altaffiliation{Both authors contributed equally}
\affiliation{School of Physical Sciences, National Institute of Science Education and Research, An OCC of Homi Bhaba National Institute (HBNI), Jatni, 752 050 Odisha, India}

\author{Ruturaj V Puranik}
\affiliation{Department of Condensed Matter Physics and Materials Science, Tata Institute of Fundamental Research, Mumbai, 400005 Maharashtra, India\looseness=-1}

\author{Vivek Dwij}
\affiliation{Department of Condensed Matter Physics and Materials Science, Tata Institute of Fundamental Research, Mumbai, 400005 Maharashtra, India\looseness=-1}

\author{Srijan Maity}
\affiliation{School of Physical Sciences, National Institute of Science Education and Research, An OCC of Homi Bhaba National Institute (HBNI), Jatni, 752 050 Odisha, India}

\author{Kanha Ram Khator}
\affiliation{School of Physical Sciences, National Institute of Science Education and Research, An OCC of Homi Bhaba National Institute (HBNI), Jatni, 752 050 Odisha, India}

\author{Subhashis Ghosh}
\affiliation{School of Physical Sciences, National Institute of Science Education and Research, An OCC of Homi Bhaba National Institute (HBNI), Jatni, 752 050 Odisha, India}

\author{Bhagwat S. Chouhan}
\affiliation{Department of Physics, Indian Institute of Technology Guwahati, Guwahati, 781 039 Assam, India}

\author{Gagan Kumar}
\affiliation{Department of Physics, Indian Institute of Technology Guwahati, Guwahati, 781 039 Assam, India}

\author{Satyaprasad P. Senanayak}
\affiliation{School of Physical Sciences, National Institute of Science Education and Research, An OCC of Homi Bhaba National Institute (HBNI), Jatni, 752 050 Odisha, India}

\author{Shriganesh Prabhu}
\affiliation{Department of Condensed Matter Physics and Materials Science, Tata Institute of Fundamental Research, Mumbai, 400005 Maharashtra, India\looseness=-1}

\author{Shovon Pal}
\altaffiliation{shovon.pal@niser.ac.in}
\affiliation{School of Physical Sciences, National Institute of Science Education and Research, An OCC of Homi Bhaba National Institute (HBNI), Jatni, 752 050 Odisha, India}

\date{\today}

\begin{abstract}
This supplementary information contains details on terahertz time-domain spectroscopy (THz-TDS) setup, fabrication of metamaterial, deposition of MAPbI$_3$, EIT metamaterial and tunability, along with theoretical treatment of the hybrid system. The contents are sectionized as:\\
\textbf{S1. Experimental Method}\\
\indent \textbf{A. Terahertz time-domain spectroscopy (THz-TDS) setup.}\\
\indent \textbf{B. Fabrication of metamaterial.}\\
\indent \textbf{C. Deposition of MAPbI$_3$.}\\
\indent \textbf{D. Choice of substrate.}\\
\textbf{S2. Strong coupling in SRR.}\\
\textbf{S3. Near-field coupling in EIT metamaterial.}\\
\textbf{S4. EIT peak amplitude variation.}\\
\textbf{S5. EIT tunability.}\\
\textbf{S6. Reproducibility of hybrid effect.}\\
\textbf{S7. Energy-level diagram for hybrid system.}\\
\textbf{S8. Theoretical model.}
\end{abstract} 

\maketitle

\section{Experimental Method}
\subsection{THz-TDS set up}
\begin{figure}[b!]
    \centering
    \includegraphics[width=0.6\linewidth]{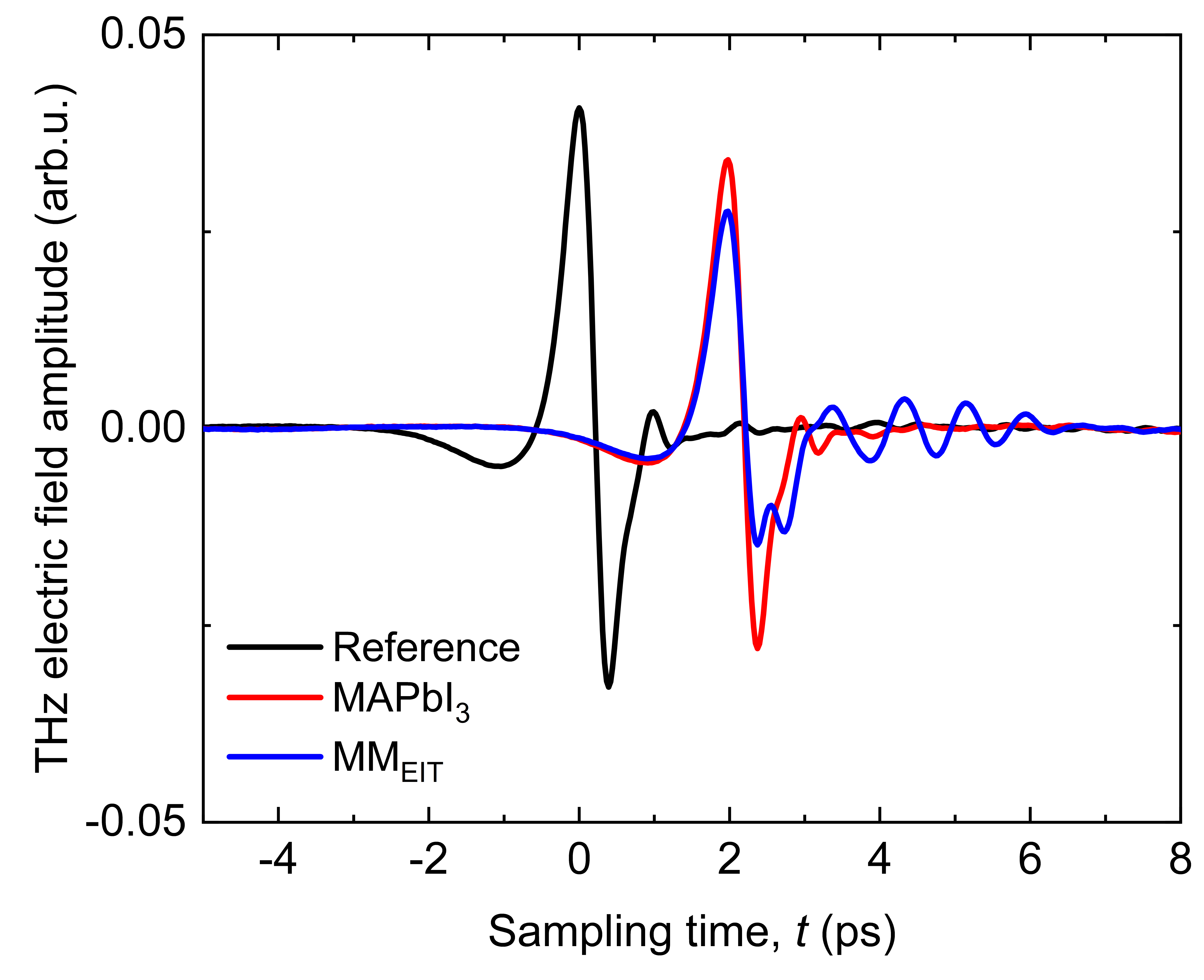}
    \caption{{Transmitted THz electric filed transients corresponding to the reference (black-curve), MAPbI$_{3}$ (red-curve) and EIT-metamaterial (blue-curve).}}
    \label{S1}
\end{figure}

We use a Ti:sapphire laser (with wavelength 800\,nm, pulse duration 10\,fs, repetition rate 80\,MHz, and pulse energy 3.75\,nJ/pulse) to generate single-cycle THz pulses from an LT-GaAs-based photoconductive antenna. While 90\% of the fundamental beam is used for the generation of THz radiation, the remaining 10\% is used as the gating beam for the free-space electro-optic sampling. A pair of parabolic mirrors is used to collimate and focus the generated THz pulses on the sample. The transmitted THz pulses are collected and focused on an optically-active 2\,mm thick ZnTe (110-cut) crystal using another pair of parabolic mirrors. The THz-induced birefringence in the detection crystal results in the rotation of polarization of the gating beam. We measure the polarization change using a quarter-wave plate, a Wollaston prism and a balanced photo-diode. All measurements are carried out in an inert nitrogen environment to eliminate water absorption lines. Figure~\ref{S1} shows a few exemplary THz time transients, namely those of reference (black-curve), and the ones transmitted through MAPbI$_{3}$ (red-curve), and the EIT-metamaterial (blue-curve) structure.

\subsection{Fabrication of metamaterial} 

\begin{figure}[b!]
    \centering
    \includegraphics[width=0.45\linewidth]{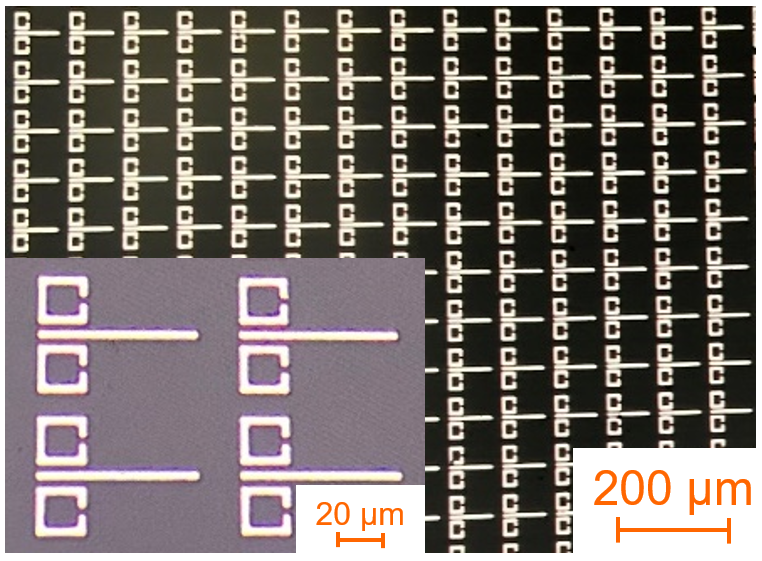}
    \caption{{Optical microscope image of the fabricated EIT structure}}
    \label{S2}
\end{figure}

Using maskless photo-lithography, we have fabricated the metamaterial structures on the $z$-cut quartz substrate (8\,mm $\times$ 8\,mm $\times$ 0.5\,mm). The $z$-cut quartz substrate provides a stable and low-loss platform that supports clean resonance features and minimizes the background absorption. For the fabrication, we used a positive photo-resist and NaOH developer. At first, the photo-resist is spin coated on the substrate with a speed of 3000\,rpm and a spinning time of 40\,s. This is followed by baking at 80\,$^\circ$C for 1\,minute. The system is then placed under the laser for the development of the structure using the maskless lithography. After the exposure, we immediately keep it in the developer solution for about 7\,s. The samples are then placed in a plasma treatment chamber to remove any organic residues that may have come from the photo-resist. The clean samples are then inserted in the thermal vapour deposition chamber for the deposition of aluminium (Al) layer of thickness 120\,nm, followed by the lift-off in N-Methyl-2-Pyrrolidone (NMP) solution. Here, the Al-layer thickness is larger than its skin depth for the spectral range of 0.5-1.6\,THz~\cite{Azad2005OptLett}. Figure~\ref{S2} shows both the unit cell and the uniformity of the fabricated metamaterial structure over an area larger the THz spot size, ensuring the homogeneous excitation of the metamaterial structure. Note that one can also use other substrate materials with different refractive indices. It would, however, change the resonance positions, affecting the obtained results qualitatively. With proper optimization of the dimensions of the metamaterial structure, we can retrieve the results. More precisely, the structure should be tuned such that after implementing the MAPbI$_{3}$ material, the resonance mode shifts and coincides with the phonon mode.

\subsection{Deposition of MAPbI$_3$}
We use the conventional solution-based spin-coating method to fabricate MAPbI$_{3}$ perovskite thin film. To fabricate the perovskite film, we prepare MAPbI$_{3}$ precursor solution by mixing the MAI (0.75\,M) and PbI$_{2}$ (0.75\,M) in N,N-Dimethylformamide (DMF) and dimethyl sulfoxide (DMSO) solvent at a volume ratio of 3:1. The precursor solution is then dropped on the $z$-cut quartz substrate for spin coating with the following conditions: spinning at 2000\,rpm for 10\,s and 6000\,rpm for 50\,s. 200\,$\mu$L of chlorobenzene (anti-solvent) is dropped onto the surface of the substrate after 40\,s from the start. The substrate is then immediately annealed on a hot-plate for 15\,minutes at 100$^\circ$\,C.

Note that MAPbI$_3$ degrades when exposed to moisture. It is thus ensured that fabrication is carefully carrier out inside nitrogen-filled glove box with both oxygen and moisture below 0.1 ppm. In addition, all the THz measurements were performed in an inert nitrogen atmosphere. Such controlled conditions ensure that MAPbI$_3$ films maintain their integrity for the entire measurement duration. Further, the experiments were carried out at room temperature and hence heat does not pose any concern on the stability of the films during our measurements. Similarly, THz light, with energies in the order of meV and energies in the order of nano-Joules, barely play any role in heating the sample and destroying the sample stability. Encapsulation of polymer coatings, however, has shown promise in improving long-time stability of the perovskite under ambient conditions~\cite{Wang2016AdvMater,Mei2014Science}. To avoid spurious signals in our experiments stemming from the polymer coating, we avoided to use of any such coatings and rather optimized to sample surrounding to minimize sample degradation.

Note that our proposed structure is well-suited for practical device integration and large-scale fabrication. This metamaterial fabrication procedure is a standard, low-cost, and scalable micro-fabrication technique compatible with wafer-scale processing. Moreover, the geometry of the metamaterial is planar, which is inherently suitable for on-chip integration with other THz components, such as waveguides or detectors. In addition, preparing the MAPbI$_3$ perovskite layer and its deposition by solution-based spin coating is widely regarded as a scalable technique. 

The resonance frequency and the coupling strength in our hybrid phonon-EIT structure are sensitive to variations in the dielectric properties of the environment. The resonance frequencies of both the rod and SRR metamaterials can be influenced by the effective refractive index of the surrounding medium. The deposition of the MAPbI$_3$ layer (with a higher refractive index compared to air) led to a significant redshift of the resonance frequencies (around 200\,GHz shift), even before activating the phonon mode. This shift arises from the change in the local electromagnetic environment, which alters the effective cavity conditions of the metamaterial resonators.

The strength of phonon-polariton coupling, quantified by the Rabi splitting, is proportional to the oscillator strength of the phonon and the local electric field enhancement around the metamaterial structures. Since the local field distribution is modified by the dielectric environment, even modest changes in environmental permittivity can influence the coupling strength. An increase in background dielectric constant generally leads to stronger field confinement, potentially enhancing the coupling strength, but it can also introduce additional absorption losses that may dampen the observed splitting.

\subsection{Choice of substrate}
\begin{figure*}[b!]
    \centering
    \includegraphics[width=0.5\linewidth]{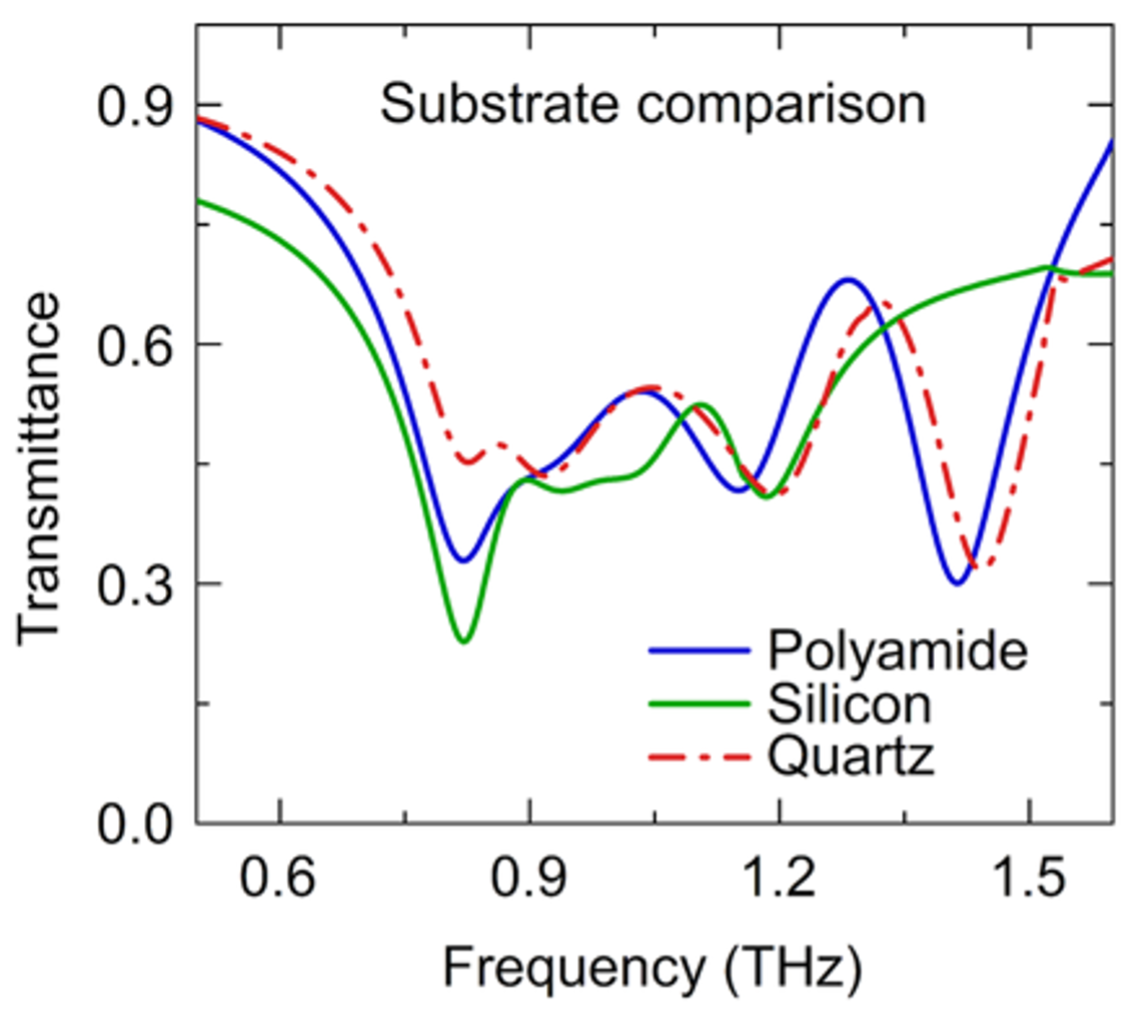}
    \caption{Simulated transmittance plots for the hybrid structure for EIT peak resonant with the phonon mode with quartz (dashed-dotted red curve) polyamide (blue-solid curve) and silicon substrate (green-solid curve.)}
    \label{S3}
\end{figure*}
Changing substrate changes the sample constellation. We have to decide on the substrate material right at the beginning since the metamaterial design depends on it. This is because different substrates have different refractive indices. Upon careful optimization of the design, one can get similar metamaterial resonance behavior, however, we found that the coupling behavior can vary depending on the substrate. To substantiate this, we performed simulations using silicon and polyamide as substrates and modified the metamaterial accordingly. The simulated results are shown in Figure~\ref{S3}. It can be seen here that the dual EIT peak features are still visible but the lower Rabi splitting in silicon restricts the frequency difference between the two peaks with lower transmission amplitude and window.

\section{Strong coupling in SRR}
\begin{figure*}[b!]
    \centering
    \includegraphics[width=0.8\linewidth]{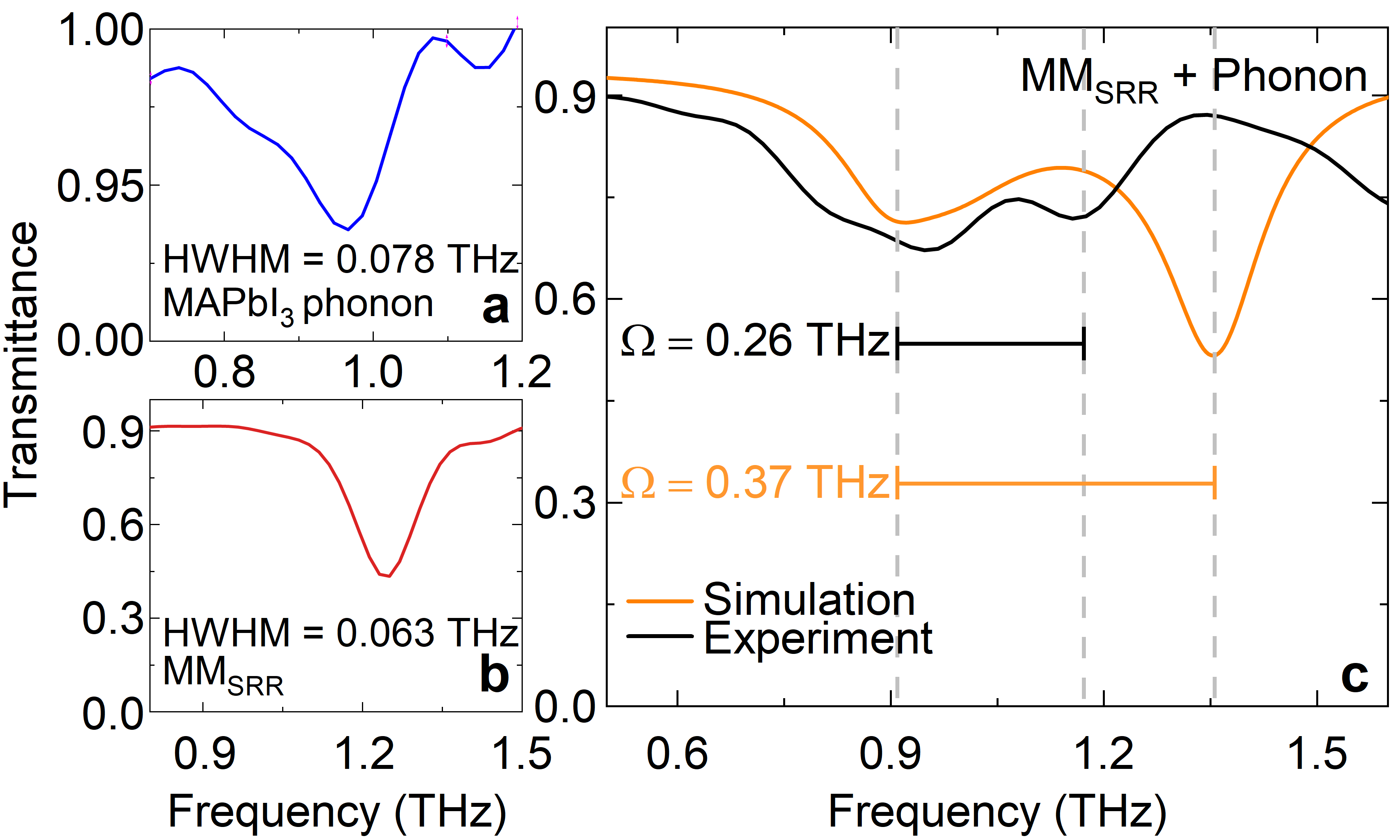}
    \caption{Experimentally measured transmittance spectra for {\bf (a)} MAPbI$_3$ phonon and {\bf (b)} SRR with corresponding HWHM of 0.078 and 0.063\,THz , respectively. Simulated and experimentally measured  transmittance spectra of the SRR in the presence of the active phonon mode of MAPbI$_{3}$ perovskite layer, showing the formation of Rabi splitting with $\Omega_{\rm R}$ = 0.37\,THz and 0.26\,THz, respectively.}
    \label{S4}
\end{figure*}
The HWHM obtained in experiment for the phonon mode and SRR is shown in Figure~\ref{S4}a and~\ref{S4}b with corresponding values of 0.078 and 0.063\,THz, respectively. Figure~\ref{S4}c presents the simulated (orange curve) and experimentally measured (black curve) transmittance spectra of the strongly-coupled SRR system. A clear splitting is observed for the phonon-polariton system. The corresponding value for energy shift due to Rabi splitting is found to be 0.26\,THz for a gap size of 3\,$\mu$m. The energy shift is substantial and allows us to reach the strong coupling regime (see manuscript for details). The coupling can be further increased by reducing the gap size in the SRR structure, however, the resonance frequency will change simultaneously. An increase in the Rabi splitting compared to our structure will enhance the transparency window and EIT amplitude in the hybrid system. The shift in magnitude and frequency is due to structural imperfections.

\section{Near-field coupling in EIT metamaterial}
\begin{figure*}[t!]
    \centering
    \includegraphics[width=0.8\linewidth]{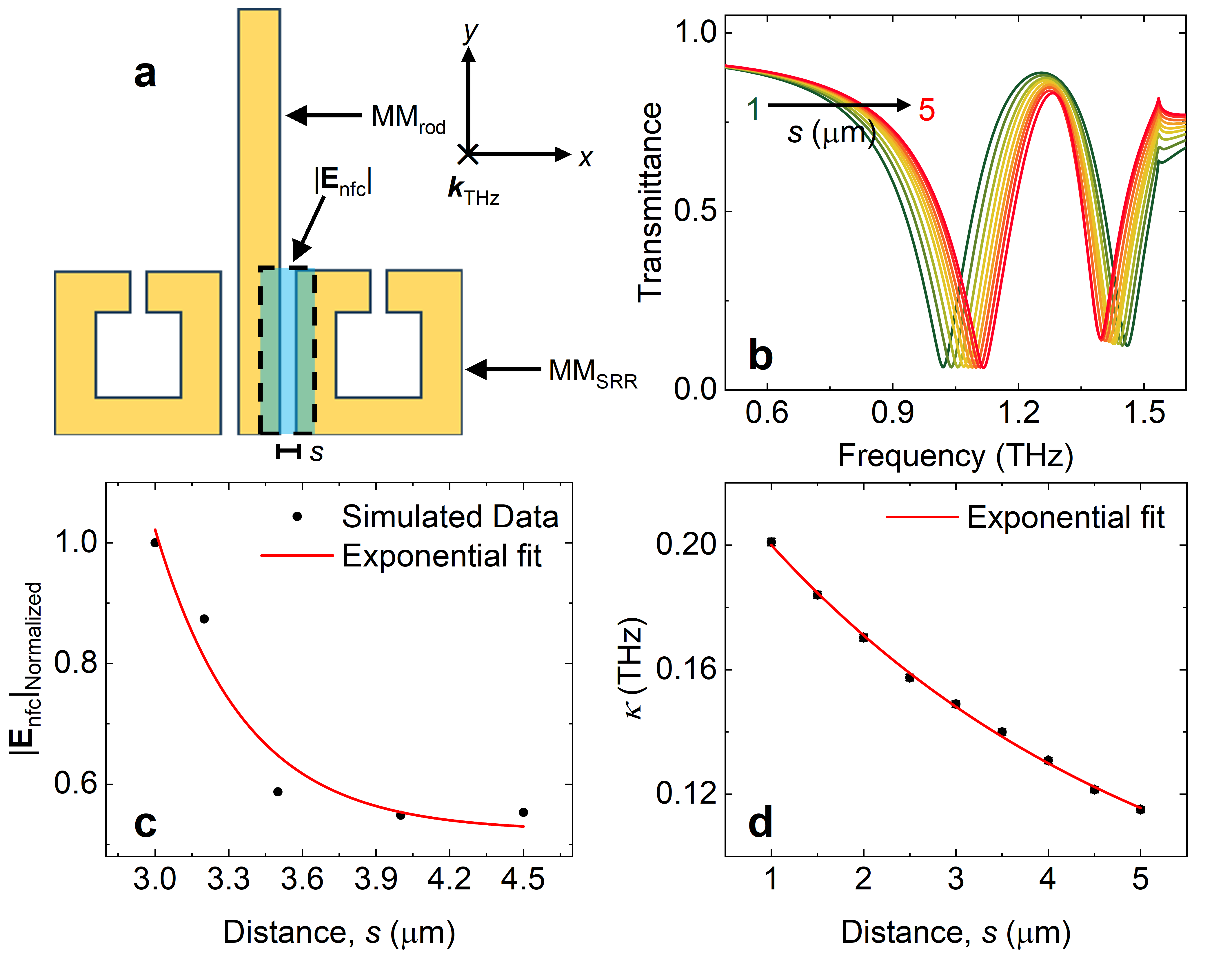}
    \caption{{\bf(a)} Schematic representation of the EIT-like metamaterial unit cell that consists of a rod resonator with a pair of split-ring resonators (SRR) on either side of the rod. The sky-blue enclosed region corresponds to the in-plane electric field distribution $|{\rm {\bf E}_{nfc}}| = \sqrt{|{\bf E}_x|^2+|{\bf E}_y|^2}$, representing the near-field coupling between the two resonators at EIT peak frequency of 1.3\,THz. {\bf(b)} Simulated transmittance spectra of the EIT-like metamaterial with varying the distance $s$ between the two resonators. The value of $s$ is changed from 1\,$\mu$m to 5\,$\mu$m in steps of 0.5\,$\mu$m. {\bf(c)} Normalized in-plane electric field distribution showing the variation of near-field coupling with changing distance between the rod and the SRR. The red solid curve represents the exponential fit to the simulated data. $|{\rm E_{nfc}}|$ is obtained by integrating the in-plane electric field over the sky-blue region for each $s$ at the EIT peak frequency of 1.3\,THz and normalized with respect to the electric field value of $s = 3$\,$\mu$m. {\bf(d)} Dependence of the coupling coefficient ($\kappa$) between the two resonators on the distance separating them. The red solid curve represents the exponential fit. Note that the value of $\kappa$ is obtained by fitting Eq.~(3) to simulated data as a function of the distance $s$ between the two metamaterial components (rod and SRR).}
    \label{S5}
\end{figure*}

Near-field coupling plays a pivotal role in the emergence of the transparency window observed in EIT-like metamaterials, arising from the destructive interference between the bright and the dark resonant modes. Unlike far-field interactions, this coupling is mediated by non-radiative evanescent fields, which dominate when the resonators are positioned in close proximity. In our design, the incident THz wave excites the bright mode of the rod, generating localized electric fields particularly concentrated at its ends. When a split-ring resonator (SRR) is placed nearby, at a separation much smaller than the resonant wavelength, it couples to the rod via this near-field mechanism. This interaction enables energy exchange between the two resonators. When the structural and resonant properties of the system are precisely tuned, this exchange leads to destructive interference, thereby opening a transparency window in the transmission spectrum. The sky-blue shaded region in Figure~\ref{S5}a highlights the spatial region where this near-field energy transfer is active.

The near-field interaction decays exponentially with the distance~\cite{Xu2020JAP} between the two resonators, marked as $s$ in Figure~\ref{S5}a. This makes the EIT response highly sensitive to the structural parameters such as the inter-resonator gap and their relative orientation. Variations in these factors can significantly impact the transparency window's bandwidth, central frequency, and spectral sharpness. This sensitivity is illustrated in the simulated transmission spectra shown in Figure~\ref{S5}b, where the inter-resonator spacing $s$ is varied from 1 to 5\,$\mu$m. A significant narrowing of bandwidth is observed for higher inter-resonator distances. We further investigate the effect of the near-field coupling in our EIT system by extracting out the simulated $|{\rm {\bf E}_{nfc}}| = \sqrt{|{\bf E}_x|^2+|{\bf E}_y|^2}$ and the coupling coefficient ($\kappa$) between the bright (rod) and dark (SRR) modes as a function of the distance $s$ between the resonators, as shown in Figures~\ref{S5}c and~\ref{S5}d. The values of $\kappa$ as shown in Figure~\ref{S5}d are obtained by fitting the transmission spectrum, represented by Eq.~(3) to the simulated transmission data generated from CST Microwave Studio, as shown in Figure~\ref{S5}b. Equation~(3) is derived by analytically solving the conventional coupled oscillator model described by Eqs.~(1) and~(2), which captures the essential dynamics of the proposed EIT-like system. From the fitting, we extract the parameters $\gamma_b$, $\gamma_d$, $\omega_b = 2\pi f_b$, $\omega_d = 2\pi f_d$ and $g$ along with $\kappa$, where a few of these are presented in Table~1 and the rest are shown in Figures~\ref{S10}a and~\ref{S10}b. We find that both the normalized in-plane electric field distribution and the coupling between the two resonators decays exponentially~\cite{Singh2009PRB,Vaswani2024OLT,Yahiaoui2017APL} with the distance between the resonators, as shown in Figures~\ref{S5}c and~\ref{S5}d. We further note that for $s = 3~\mu$m, the coupling $\kappa^2$ between the resonators is greater than $(\gamma_b\gamma_d)$~\cite{Wan2015PLA,Singh2014APL}, which indicates that the two resonators are strongly near-field coupled for the energy exchange to happen that has already been established in an earlier report~\cite{Liu2012APL}.

\begin{table}[t!]
\centering
\caption{\normalsize {Parameters, such as the bright and dark mode damping parameters ($\gamma_a$ and $\gamma_b$) and the coupling constant $\kappa$ obtained from fitting Eq.~(3) to the simulated data. Note for all chosen distances $s$, $\kappa^2 > \gamma_b\gamma_d$, validating the strong near-field coupling~\cite{Wan2015PLA,Singh2014APL}.}}
\vspace{3pt}
\renewcommand{\arraystretch}{1.0}
\setlength{\tabcolsep}{5pt}
\begin{tabular}{>{\centering\arraybackslash}p{1.8cm} *{5}{>{\centering\arraybackslash}p{2.0cm}} >{\centering\arraybackslash}p{1.8cm}}
\hline
\toprule
\textbf{Distance} & $\boldsymbol{\gamma_b}$ & $\boldsymbol{\gamma_d}$ & $\boldsymbol{\kappa}$ & $\boldsymbol{\gamma_1\gamma_2}$ & $\boldsymbol{\kappa^2}$ \\
$s$ ($\mu$m) & (THz) & (THz) & (THz) & (THz$^2$) & (THz$^2$) \\ \hline \hline
\midrule
1.0 & $0.119$ & $0.087$ & $0.201$ & $0.010$ & $0.040$ \\  \hline
1.5 & $0.126$ & $0.077$ & $0.184$ & $0.010$ & $0.034$ \\  \hline
2.0 & $0.134$ & $0.070$ & $0.170$ & $0.009$ & $0.029$ \\  \hline
2.5 & $0.137$ & $0.063$ & $0.157$ & $0.009$ & $0.025$ \\  \hline
3.0 & $0.142$ & $0.059$ & $0.149$ & $0.008$ & $0.022$ \\  \hline
3.5 & $0.147$ & $0.054$ & $0.140$ & $0.008$ & $0.020$ \\  \hline
4.0 & $0.150$ & $0.050$ & $0.131$ & $0.008$ & $0.017$ \\  \hline
4.5 & $0.153$ & $0.046$ & $0.121$ & $0.007$ & $0.015$ \\  \hline
5.0 & $0.157$ & $0.042$ & $0.115$ & $0.007$ & $0.013$ \\  \hline
\bottomrule
\end{tabular}
\end{table}

\section{EIT peak amplitude variation}
\begin{figure*}[t!]
    \centering
    \includegraphics[width=0.55\linewidth]{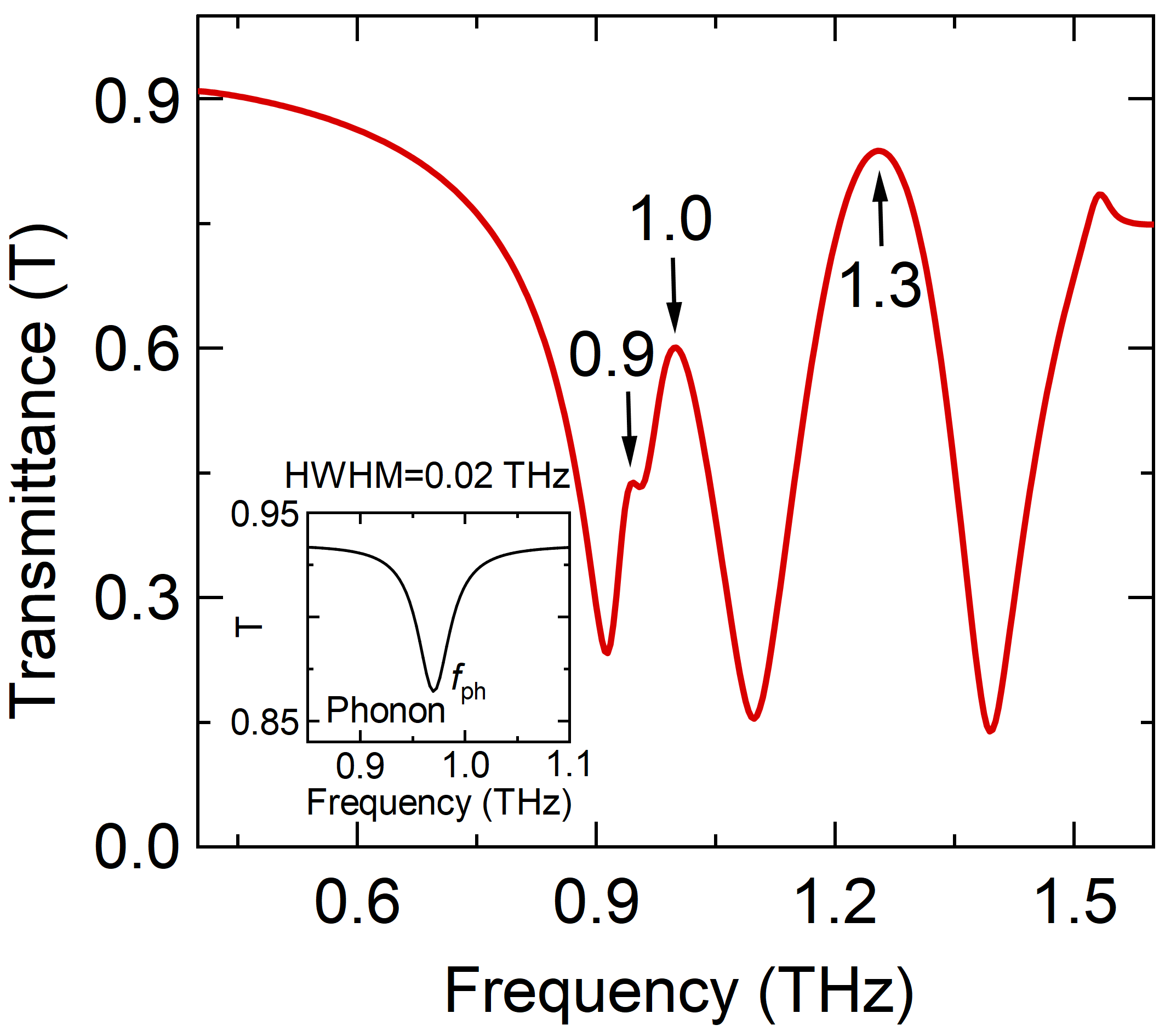}
    \caption{Simulated transmittance spectra of phonon-EIT hybrid metamaterial structure with a sharp phonon mode at $f_{ph}=0.97$\,THz shown in the inset. The maxima obtained show much stronger transmittance specially for 1.3\,THz EIT-like peak.}
    \label{S6}
\end{figure*}
One of the key challenges faced in the hybrid structure that we demonstrated is the limited amplitude of the maxima. This can actually be transcended by choosing a material with a sharper phonon mode. A sharper phonon increases the transmittance dips obtained for individual Rabi splitting which thereby increases the EIT-like peak amplitude. To demonstrate this, we simulated an alternate hypothetical phonon mode with a considerably smaller half-width at half maximum (HWHM) = 0.02\,THz as shown by the inset of Figures~\ref{S6}. Such a phonon, gives us transmittance as high as the original EIT-like peak for the EIT-like maximum at 1.3 \,THz. Low temperature measurements would further improve the phonon sharpness. Stronger phonon modes found in materials such as PbTe~\cite{baydin2025arxiv}, can help achieve stronger coupling regimes such as ultra-strong or even deep-strong that can induce further splitting in the hybrid system.

\section{EIT tunability}
\begin{figure*}[b!]
    \centering
    \includegraphics[width=0.75\linewidth]{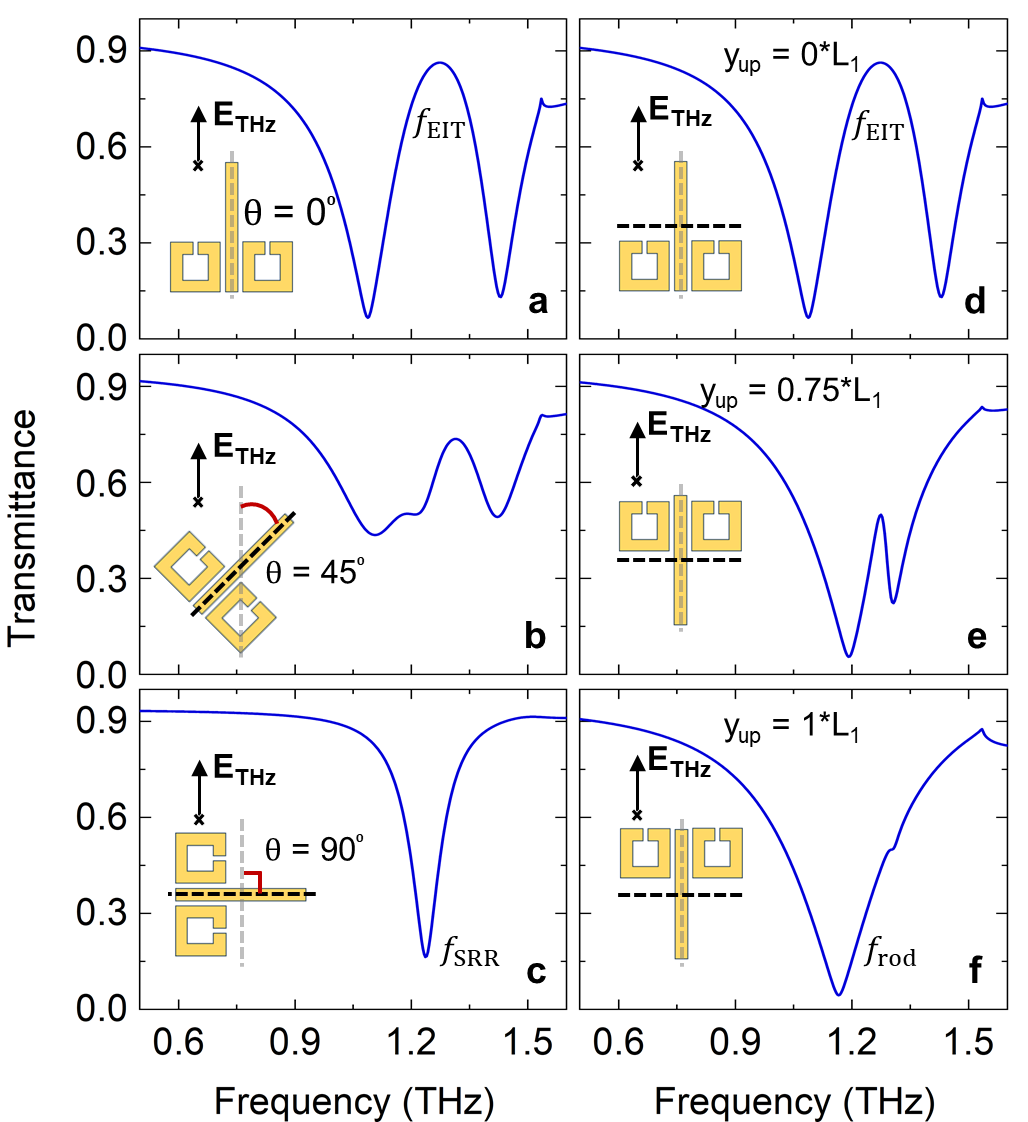}
    \caption{Simulated transmittance spectra of EIT metamaterial structure with {\bf (a-c)} rotational and {\bf (d-f)} translation tunability. $\theta$ is the angle of rotation of the metamaterial and $y_{\rm up}$ is the magnitude of upward translation of the SRR along the length of the rod in $\mu$m. Here, the reference point for translation is the center of the SRR at $y_{\rm up}=0$. We define $L_1=L-l$.}
    \label{S7}
\end{figure*}

The rod-SRR EIT structure was specifically chosen for this study due to the rotational and translational tunability possible in this geometry. Such a tunability enables us to explore the dynamic switching from strong coupling to the hybrid phonon-EIT system. As described in the main text, the rod and SRR are directly excited by orthogonal polarizations of the incident electromagnetic wave. This means that rotating the structure would allow us to deactivate the resonance of the rod and simultaneously directly excite the SRR resonance. The de-excitation of the rod with rotation prevents the hybridization of the bright-dark states thereby destroying the EIT nature and transitioning to the SRR resonance. This can be seen in the simulated transmittance spectra in Figures~\ref{S7}a-c, wherein a clear transition of the EIT nature to a LC resonance (of SRR) can be mapped, simply by rotating our metamaterial. 

Another degree of tunability in this structure comes from translating the SRR along the length of the rod. This EIT structure in particular has dual excitation pathways, namely electric and magnetic. At $y_{\rm up}=0*L_1$, the current induced in the SRR due to the electric and magnetic fields is in phase and thus constructively interferes to give the LC resonance. However, as the SRR is translated upwards, the current due to the electric excitation switches direction. This leads to a destructive interference between the excitation pathways consequently destroying the LC resonance of the SRR and hence the EIT nature, transitioning to the dipole resonance of the rod as shown in Figure~\ref{S7}d-f. To put it simply, as the SRR is translated upwards, the coupling mechanism which was initially capacitative in nature, acquires a more inductive behavior and has been demonstrated before~\cite{Liu2012APL}.

\begin{figure*}[h!]
    \centering
    \includegraphics[width=1\linewidth]{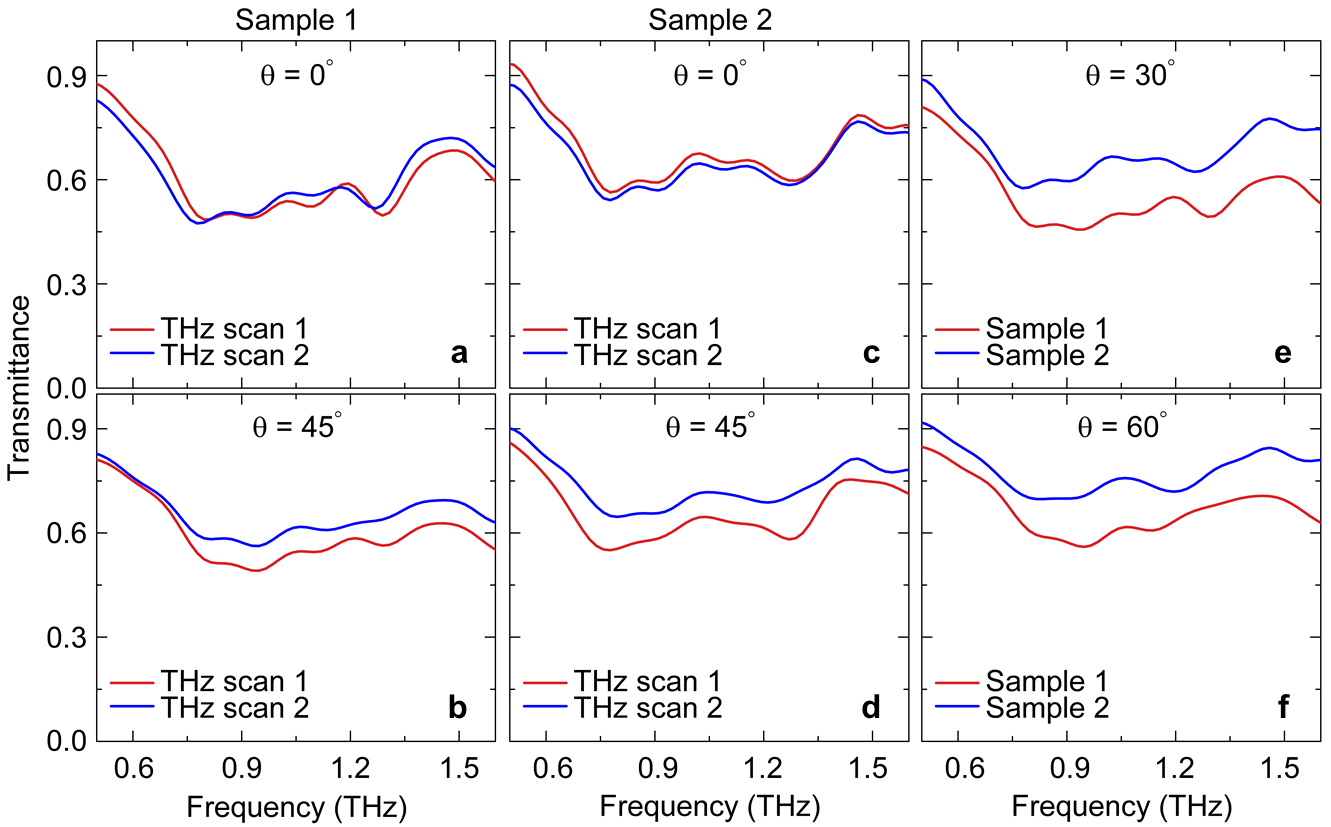}
    \caption{{\bf (a)} Reproducibility of experimentally obtained transmittance for metamaterial sample 1 at {\bf (a)} $0^{\circ}$ and {\bf (b)} $90^{\circ}$. Reproducibility of experimentally obtained transmittance for metamaterial sample 2 at {\bf (c)} $\rm\theta=0^\circ$ and {\bf (d)} $90^{\circ}$. {\bf (e)} and {\bf (f)} show the experimentally measured transmittance for $\rm\theta=30^{\circ}$ and $60^{\circ}$ for sample 1 and 2. $\rm\theta=0^{\circ}$ implies THz pulse is polarized along the length of the rod. The metamaterial is rotated by $\rm\theta$ to effectively rotate the polarization angle of the input THz wave by the same angle.}
    \label{S8}
\end{figure*}

\section{Reproducibility of hybrid effect}
The robustness and reproducibility of the tuning effect was verified both by repeating the measurement for the same sample and also using a second sample with a slightly shifted EIT peak at 1.26 THz labeled as sample 2 in Figure 1. For sample 1, for $\rm\theta=0^\circ$ and $45^{\circ}$, both the scans show good agreement as shown in Figure~\ref{S8}a and~\ref{S8}b.

Similarly, for sample 2, both angles show excellent agreement, with a slight difference in transmittance; however, the features are nicely reproducible (Figure~\ref{S8}c and~\ref{S8}d). Hence, the reproducibility of the effect is verified both by multiple scans and different metamaterial samples. Furthermore, we investigated other angles such as $\rm\theta=30^\circ$ and $60^\circ$ for both the samples given in Figure~\ref{S8}e and~\ref{S8}f.

\section{Energy-level diagram for hybrid system}
\begin{figure*}[t!]
    \centering
    \includegraphics[width=\linewidth]{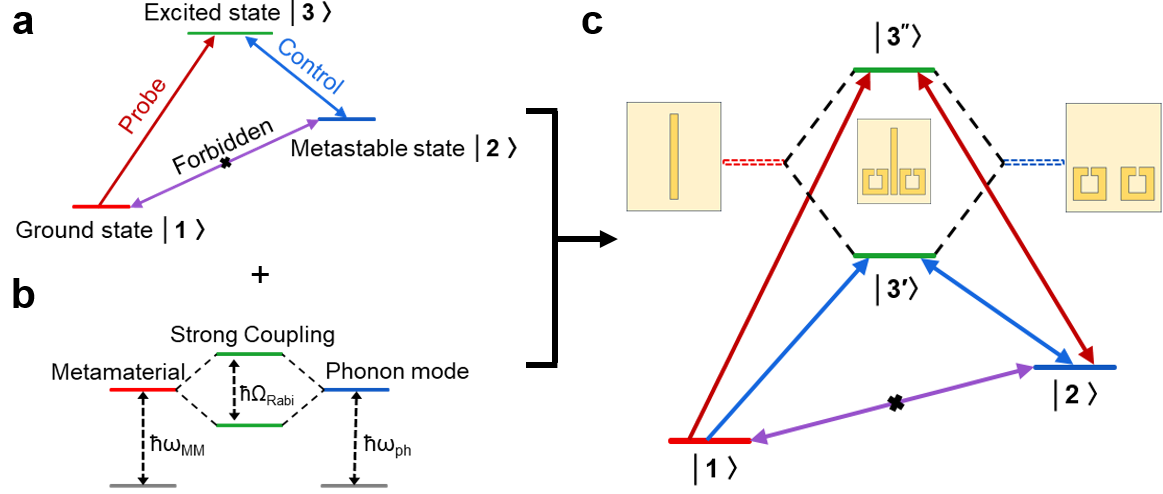}
    \caption{{\bf (a)} Energy-level diagram for an EIT system. The transition from $\ket{1}$ to $\ket{2}$ and vice versa is dipole forbidden. The system is made transparent to the resonant lasing transition from $\ket{1}$ to $\ket{3}$ transition by carefully tuning a control field to achieve destructive interference of the probability amplitudes. {\bf (b)} Energy-level diagram for the Rabi splitting in a phonon-polariton system. {\bf (c)} Energy-level diagram to couple strong coupling with the EIT effect. The rod and SRR states are represented using the red and blue energy states, respectively. Rabi splitting due to phonon creates $\ket{3'}$ and $\ket{3''}$, marked by the green energy levels. Two distinct excitation pathways indicated by red and blue arrows suggest the presence of dual EIT nature emerging in the EIT-phonon system.}
    \label{S9}
\end{figure*}

EIT in metamaterials is an established classical analog to the quantum mechanical EIT effect. To take this analogy a step forward, we incorporate the strong coupling effect into an EIT level diagram. Figure~\ref{S9}a shows the standard energy level diagram for an EIT system. The probability amplitude of the electrons excited by the probe and control field have a $\pi$ phase difference that leads to the net destructive interference in $\ket{3}$. When precisely tuned, the probability amplitude in $\ket{3}$ can be made zero, thereby inducing a transparency window. For Rabi splitting, on the other hand, an energy splitting is observed for a strongly-coupled system for sufficiently close resonances, Figure~\ref{S9}b. We now intend to combine these systems.

Previous studies have discussed the modified energy level system for a dual-EIT-based system~\cite{Xu2016AdvOptMater,Chen2020OptMat,Sarkar2020JOpt}. In our case, however, additional energy levels come into play to account for the strong coupling effect. To do this, we refer to our order of excitation, discussed in the main text. The rod and the SRR undergo individual Rabi splitting to give rise to a pair of lower- and upper-polaritonic states. The significant overlap of the higher and lower resonance dips for the rod and SRR allow us to condense the two pair of states to single lower-polaritonic and upper-polaritonic states denoted by $\ket{3'}$ and $\ket{3''}$ respectively, as shown in Figure~\ref{S9}c. This provides dual excitation pathways for EIT to take place at two distinct frequencies modulated by the amount of splitting, as shown by the red and blue arrows in Figure~\ref{S9}c.

\section{Theoretical Model}
\begin{figure*}[t!]
    \centering
    \includegraphics[width=0.9\linewidth]{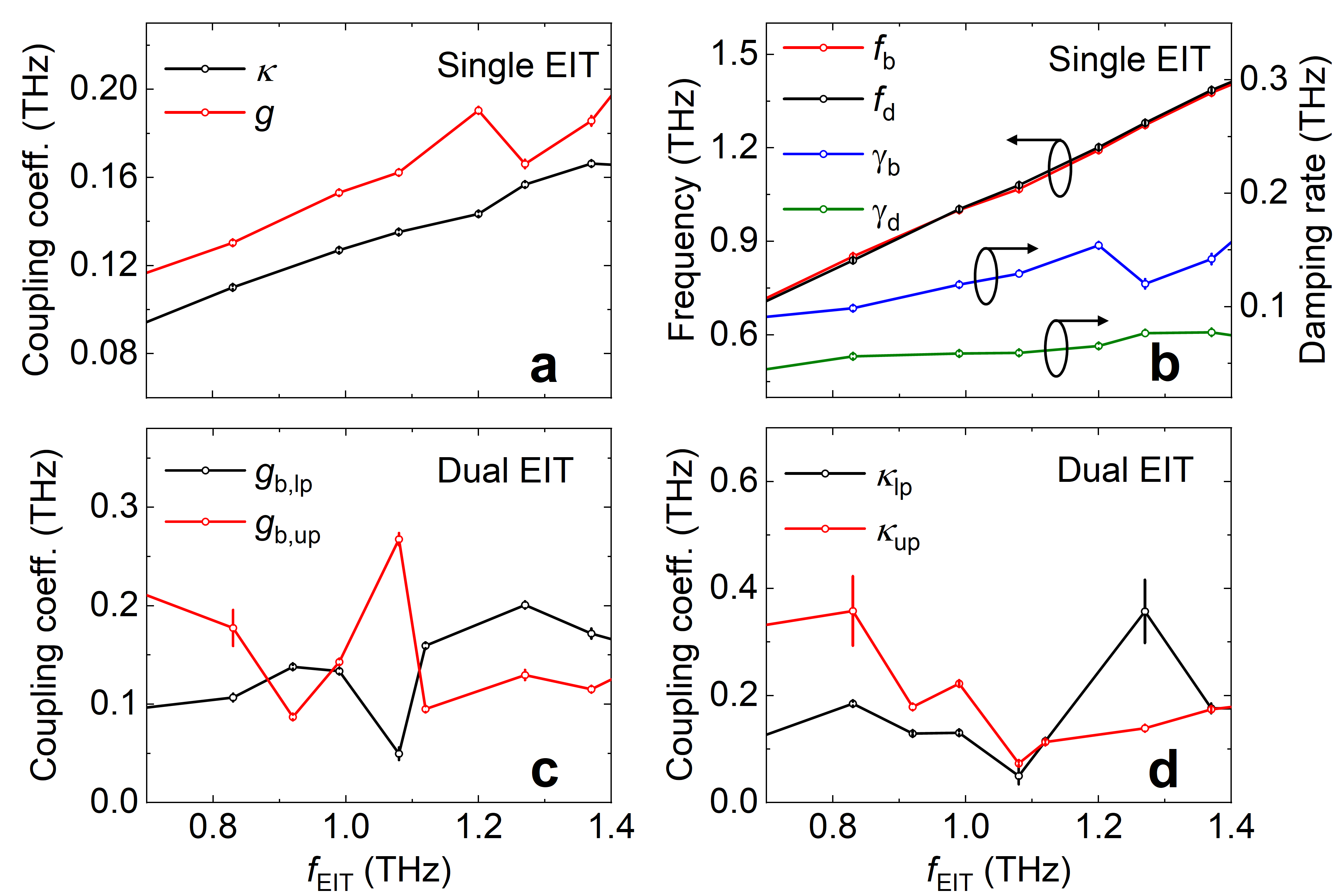}
    \caption{Extracted coupling coefficient parameters ($\kappa$, $g$, $f$ and $\gamma$) as a function of $f_{\rm EIT}$ for single {\bf (a-b)} and dual {\bf(c-d)} EIT. {\bf(a)} $\kappa$ is the coupling coefficient between the bright and dark modes of the single EIT-like system. $g$ is the coupling coefficient demonstrating the coupling strength of the bright mode with the incident electromagnetic wave. {\bf (b)} $f_b$ and $f_d$ are the resonance frequencies of the bright and dark modes, respectively. $\gamma_b$ and $\gamma_d$ are the damping rates of the bright and dark modes, respectively. {\bf(c)} $g_{\rm{b,lp}}$ and $g_{\rm{b,up}}$ are the coupling coefficients demonstrating the coupling strength of the bright mode of the lower and upper frequency with the incident electromagnetic wave, respectively. Here, this bright mode corresponds to the polaritonic state at the lower and the upper frequency. {\bf(d)} $\kappa_{\rm lp}$ and $\kappa_{\rm up}$ are the coupling coefficients between the bright and dark modes of the hybrid system at the lower and the upper frequency, respectively.}
    \label{S10}
\end{figure*}

We use two-coupled, three-level system to model our hybrid phonon EIT-like system. We start with the EIT-like structure, where the rod interacts directly with the input electromagnetic field, allowing an electric-dipole transition, and serves as the bright mode. In addition, the SRR pair subsequently gets excited through the $x$-component of the THz electric field induced by the rod and acts as the dark mode. In this scenario, the field induced by the dipole excitation of the rod destructively interferes with the capacitance gap of the SRR, thereby opening up a transparency window. While  the extended coupled oscillator (ECO) model provides a more general description of radiative interference~\cite{Lovera2013ACSNano}, we adopt the conventional coupled oscillator (CCO) model for our EIT-like system. This adaptation is appropriate because, in our EIT-like structure, only the bright mode (i.e., the rod) couples significantly to the external electromagnetic field. In contrast, the dark mode (i.e., the SRRs) is effectively shielded from direct excitation. The simplified CCO model accurately captures the key features of the destructive interference and the transparency without requiring additional complexity. Within the framework of CCO, the destructive interference can be modeled using the following equations of motion for the bright and dark oscillators:
\begin{align}
    \ddot{x}_b(t) + \gamma_b\dot{x}_b(t) + \omega_b^2x_b(t) + \kappa^2x_d(t)  &= gE_0, \\
    \ddot{x}_d(t) + \gamma_d\dot{x}_d(t) + \omega_d^2x_d(t) + \kappa^2x_b(t) &= 0.
\end{align}
Here, $x_{\rm b}$ and $x_{\rm d}$ are the amplitudes of the bright (rod) and dark (SRR) mode resonators. $\omega_{\rm b}$ and $\omega_{\rm d}$ are the resonance frequencies and $\gamma_{\rm b}$ and $\gamma_{\rm d}$ are the damping rates of the bright and dark resonators, respectively. $g$ is the coupling coefficient that mediates the coupling of the bright mode with the incident electromagnetic field and $\kappa$ is the coupling coefficient between the bright mode and dark mode resonators. The transmission for the proposed EIT-like system can be expressed using the equation~\cite{Ling2018Nanoscale}:
\begin{align}
    {\rm T} = 1 - \left|\frac{g(\omega - \omega_{\rm d} + \textit{i}\gamma_{\rm d})}{(\omega - \omega_{\rm b} + \textit{i}\gamma_{\rm b})(\omega - \omega_{\rm d} + \textit{i}\gamma_{\rm d})+\kappa^2}\right|^2.
\end{align}
The simulated EIT-like spectra is fitted with Eq.~3 and is shown in Figure~1d of the main manuscript. Figure~\ref{S10}a shows the fitting parameters, namely the coupling coefficients, $\kappa$ and $g$, as a function $f_{\rm EIT}$ for the single EIT-like system. We find that the coupling coefficients increase with the $f_{\rm EIT}$. Note that for the designed $f_{\rm EIT} = 1.3$\,THz, the resonator-resonator coupling strength is sufficient for it to be in the strong near-field coupling regime. Figure~\ref{S10}b presents the fitting parameters, including the resonance frequencies and damping rates of the bright and dark modes, respectively. While both the frequencies of the bright and dark modes increase with the $f_{\rm EIT}$, their damping rates remain almost constant.

To incorporate the Rabi splitting effect, we note that in strong coupling regime the eigen-frequencies of the lower (lp) and the upper (up) polaritons are given by,
\begin{align}
    \omega_{\rm {lp,up}}=\omega_0\pm g_{\rm {rabi}},
\end{align}
where $g_{\rm {rabi}}$ is the coupling strength with the Rabi splitting $\Omega_{\rm R} = 2g$ and $\omega_0$ being the central frequency of the TO-phonon in MAPI. As proposed in our interpretation, for the emergence of the three maxima, shown in the main manuscript, both the bright and dark modes undergo Rabi splitting to give $\omega_{\rm b,lp}$, $\omega_{\rm b,up}$ and $\omega_{\rm d,lp}$, $\omega_{\rm d,up}$, respectively. Similarly, the new resonances resulting from the splitting have different decay rates $\gamma_{\rm b,lp}$, $\gamma_{\rm b,up}$ and $\gamma_{\rm d,lp}$, $\gamma_{\rm d,up}$ for the bright and dark mode. The bright mode now couples to the incident field with two distinct coupling coefficients $g_{\rm b,lp}$ and $g_{\rm b,up}$. The pair of Rabi dips obtained (shown in Figure~2c and~2d of the main manuscript), give rise to a pair of bright-dark resonances that destructively interfere to give a dual EIT effect. This means that we now have a pair of coupling coefficients $\kappa_{\rm lp}$ and $\kappa_{\rm up}$, between the bright and dark modes of the hybrid system corresponding to the lower-polaritonic-EIT and upper-polaritonic-EIT peaks. Hence, in Eq.~3, $\omega_{\rm d}$ is replaced by $\omega_{\rm d,up,lp}$ and $\omega_{\rm b}$ by $\omega_{\rm b,up,lp}$. The two resonances, give rise to two separate transparency windows. In such a case, the net transmittance is then given by the superposition of the obtained transparency windows, i.e.,
\begin{eqnarray}
    {\rm T_{dEIT}} = 1 - \left|\frac{g_{\rm b,lp}(\omega - \omega_{\rm d,lp} + \textit{i}\gamma_{\rm d,lp})}{(\omega - \omega_{\rm b,lp} + \textit{i}\gamma_{\rm b,lp})(\omega - \omega_{\rm d,lp} + \textit{i}\gamma_{\rm d,lp})+\kappa_{\rm lp}^2}\right|^2 \nonumber \\
    -\left|\frac{g_{\rm b,up}(\omega - \omega_{\rm d,up} + \textit{i}\gamma_{\rm d,up})}{(\omega - \omega_{\rm b,up} + \textit{i}\gamma_{\rm b,up})(\omega - \omega_{\rm d,up} + \textit{i}\gamma_{\rm d,up})+\kappa_{\rm up}^2}\right|^2
\end{eqnarray}
This model fits quite nicely to the simulated data as shown in Figure.~3c of the main manuscript. All the fitting parameters for the hybrid system are extracted by fitting Eq.~5 to the simulated data. Figures~\ref{S10}c and~\ref{S10}d show the coupling coefficients demonstrating the coupling strength of the bright mode with the incident electromagnetic wave and between the bright and dark modes of the hybrid system, respectively.